\documentclass[12pt,letterpaper]{emulateapj}
\usepackage{rotating}
\usepackage[stable]{footmisc}
\shorttitle{Heating and Cooling in NGC1097}
\shortauthors{Beir\~ao et al.}
\begin{document}
 
\title{A Study of Heating and Cooling of the ISM in NGC 1097 with Herschel-PACS and Spitzer-IRS}

\author{P. Beir\~ao\altaffilmark{1},
       L. Armus\altaffilmark{1},        
       G. Helou\altaffilmark{2},
        P. N. Appleton\altaffilmark{3}, 
       J.-D. T. Smith \altaffilmark{5},
         K. V. Croxall\altaffilmark{5},
        E. J. Murphy\altaffilmark{4}, 
      D. A. Dale\altaffilmark{6},
      B. T. Draine\altaffilmark{7},
         M. G. Wolfire\altaffilmark{8},
         K. M. Sandstrom\altaffilmark{9,22},
         G. Aniano\altaffilmark{7},
        A. D. Bolatto\altaffilmark{8},
          B. Groves\altaffilmark{9},
        B. R. Brandl\altaffilmark{10},
        E. Schinnerer\altaffilmark{9},
         A. F. Crocker\altaffilmark{13},
         J. L. Hinz\altaffilmark{11},
           H.-W. Rix\altaffilmark{9},
        R. C. Kennicutt\altaffilmark{12},
       D. Calzetti\altaffilmark{13},
        A. Gil de Paz\altaffilmark{14},
     G. Dumas\altaffilmark{9},   
      M. Galametz\altaffilmark{12},
      K. D. Gordon\altaffilmark{15},
      C.-N. Hao\altaffilmark{16},
       B. Johnson\altaffilmark{20},
     J. Koda\altaffilmark{17},
       O. Krause\altaffilmark{9},
         T. van der Laan\altaffilmark{9},
          A. K. Leroy\altaffilmark{18,23},
          Y. Li\altaffilmark{13},
     S. E. Meidt\altaffilmark{9},
     J. D. Meyer\altaffilmark{17},
        N. Rahman\altaffilmark{8},
                   H. Roussel\altaffilmark{20},
                M. Sauvage\altaffilmark{20},
                  S. Srinivasan\altaffilmark{20},
                  L. Vigroux\altaffilmark{21},
                  F. Walter\altaffilmark{9},
          B. E. Warren\altaffilmark{19}
}

\altaffiltext{1}{Spitzer Science Center, California Institute of Technology, Pasadena, CA 91125}
\altaffiltext{2}{Infrared Processing and Analysis Center, California Institute of
Technology, Pasadena, CA 91125}
\altaffiltext{3}{NASA Herschel Science Center, California Institute of
Technology, Pasadena, CA 91125}
\altaffiltext{4}{Carnegie Observatories, Pasadena, CA 91101, USA}
\altaffiltext{5}{Department of Physics and Astronomy, Mail Drop 111, University
              of Toledo, 2801 West Bancroft Street, Toledo, OH 43606, USA}
\altaffiltext{6}{Department of Physics \& Astronomy, University of Wyoming,
              Laramie, WY 82071, USA}           
\altaffiltext{7}{Department of Astrophysical Sciences, Princeton University,
              Princeton, NJ 08544, USA}
\altaffiltext{8}{Department of Astronomy, University of Maryland, College Park,
              MD 20742, USA}
\altaffiltext{9}{Max-Planck-Institut f\"ur Astronomie, K\"onigstuhl 17, 69117
              Heidelberg, Germany}
\altaffiltext{10}{Leiden Observatory, Leiden University, P.O. Box 9513, 2300 RA
              Leiden, The Netherlands}
\altaffiltext{11}{Steward Observatory, University of Arizona, Tucson, AZ 85721,
              USA} 
\altaffiltext{12}{Institute of Astronomy, University of Cambridge, Madingley Road,
              Cambridge CB3 0HA, UK}
\altaffiltext{13}{Department of Astronomy, University of Massachusetts, Amherst,
              MA 01003, USA}
\altaffiltext{14}{Departamento de Astrofisica, Facultad de Ciencias Fisicas,
              Universidad Complutense Madrid, Ciudad Universitaria, Madrid, E-28040, Spain}
\altaffiltext{15}{Space Telescope Science Institute, MD 21218, USA}
\altaffiltext{16}{Tianjin Astrophysics Center, Tianjin Normal University, 
              Tianjin 300387, China}
\altaffiltext{17}{Department of Physics and Astronomy, SUNY Stony Brook, Stony
              Brook, NY 11794-3800, USA}
\altaffiltext{18}{National Radio Astronomy Observatory, 520 Edgemont Road, 
              Charlottesville, VA 22903, USA}
\altaffiltext{19}{ICRAR, M468, University of Western Australia, 35 Stirling Hwy, Crawley, WA, 6009, Australia} 
\altaffiltext{20}{Institut d'Astrophysique de Paris, UMR7095 CNRS
              Universit\'e Pierre \& Marie Curie, 98 bis boulevard Arago, 75014 Paris, France}
\altaffiltext{21}{CEA/DSM/DAPNIA/Service d'Astrophysique, UMR AIM, CE Saclay,
              91191 Gif sur Yvette Cedex} 
\altaffiltext{22}{Marie Curie fellow}
\altaffiltext{23}{Hubble Fellow}   
         
\email{pedro@ipac.caltech.edu}

\begin{abstract}
NGC 1097 is a nearby Seyfert 1 galaxy with a bright circumnuclear starburst ring, a strong large-scale bar and an active nucleus. We present a detailed study of the spatial variation of the far infrared (FIR) [CII]$158\mu$m and [OI]$63\mu$m lines and mid-infrared $H_2$ emission lines as tracers of gas cooling, and of the polycyclic aromatic hydrocarbon (PAH) bands as tracers of the photoelectric heating, using Herschel-PACS, and Spitzer-IRS infrared spectral maps. We focus on the nucleus and the ring, and two star forming regions (Enuc N and Enuc S). We estimated a photoelectric gas heating efficiency ([CII]$158\mu$m+[OI]$63\mu$m)/PAH in the ring about 50\% lower than in Enuc N and S. The average 11.3/7.7$\mu$m PAH ratio is also lower in the ring, which may suggest a larger fraction of ionized PAHs, but no clear correlation with [CII]158$\mu$m/PAH($5.5-14\mu$m) is found. PAHs in the ring are responsible for a factor of two more [CII]$158\mu$m and [OI]$63\mu$m emission per unit mass than PAHs in the Enuc S. SED modeling indicates that at most 25\% of the FIR power in the ring and Enuc S can come from high intensity photodissociation regions (PDRs), in which case $G_0 \sim10^{2.3}$ and $n_H \sim 10^{3.5}$ cm$^{-3}$ in the ring. For these values of $G_0$ and $n_H$ PDR models cannot reproduce the observed $H_2$ emission. Much of the the $H_2$ emission in the starburst ring could come from warm regions in the diffuse ISM that are heated by turbulent dissipation or shocks. 
\end{abstract}



\section{Introduction}

The most important heating process in the neutral ISM is expected to be the photoelectric effect working on polycyclic aromatic hydrocarbons (PAHs) and small dust grains \citep{watson72, tielens85, wolfire95, holl99}. Incident far-ultraviolet photons have energies high enough to eject electrons from dust grains (h$\nu \gtrsim 6$ eV). They heat the gas via these photoelectrons, with a typical energy efficiency of $0.1\%-1\%$. The efficiency is determined mostly by the ratio of UV radiation field to gas density ($G_0/n$) which sets the charge of the grains \citep{bakes94,weing01}. The heated gas cools primarily via fine-structure lines such as [CII]$158\mu$m and [OI]$63\mu$m, with [CII]$158\mu$m being dominant at lower densities and temperatures.
The photoelectric effect on dust thus couples the far-ultraviolet radiation field to the interstellar gas heating outside of H II regions. Such regions where the dust photoelectric effect
dominate gas heating and/or chemistry are called photodissociation regions \citep{holl99}. Rotational transition lines of $H_2$ are also produced in photodissociation regions (PDRs), at the interface between the molecular gas and the atomic gas.
The heating efficiency of the photoelectric effect is typically measured as the photoelectric yield per heating luminosity, using the flux ratio of the [CII]$158\mu$m and/or [OI]$63\mu$m to the far-infrared (FIR) luminosity. A further discussion on the definition of the photoelectric heating efficiency can be found in Section \ref{discgas}.

Observations of the FIR cooling lines in representative samples of local star-forming galaxies were pioneered with the Kuiper
Airborne Observatory, but were first studied in local star-forming galaxies with ISO \citep{malhotra97, malhotra01, helou01, contursi02}. These studies clearly showed that the flux ratio [CII]$158\mu$m/L(FIR) decreases by more than an order of magnitude (from 0.004 to less than 0.0004) for galaxies with high $L(IR)/L(B)$ and/or warm dust temperatures while the [CII]$158\mu$m/PAH ratios show no such trends with dust temperature or luminosity. This means that the role of PAHs on gas heating decreases with $L(FIR)/L(B)$, which could be attributed to PAH ionization, or that [CII] becomes a less important cooling channel, resulting in more cooling via [OI]$63\mu$m. 
In ``active" regions with high FIR luminosity, [OI]$63\mu$m becomes a more important cooling mechanism than [CII]$158\mu$m \citep{malhotra01}.
The mechanism responsible for heating these ``active'' regions is uncertain, but shock heating is one candidate, and energy injection from deeply embedded star formation in dense regions is another. Models of the integrated emission of [OI]$63\mu$m, [CII]$158\mu$m, $H_2$ vibrational modes and other emission lines from Galactic and extragalactic star-forming regions have been used to diagnose these mechanisms \citep[e.g.][]{holl89,kaufman06}. Unfortunately the low spatial resolution of ISO at long wavelengths and
the modest samples of sources detected in multiple lines make direct comparisons to models extremely difficult.
With Herschel-PACS and Spitzer-IRS, there is now the opportunity to use the PAH bands and the main infrared cooling lines to study the variations of the photoelectric heating efficiency and diagnose the main heating mechanisms of the ISM in an environment with a wide range of ISM phases.

\begin{deluxetable*}{lcccc}
\tablecaption{Exposure times}
\tablewidth{0pt}
\tablehead{
\colhead{Region} & \colhead{SL} & \colhead{LL} & \colhead{SH} & \colhead{LH}\\
 & \colhead{($5.5 - 14.5\mu$m)} & \colhead{($14 - 38\mu$m)} & \colhead{($10.5 - 19.5\mu$m)} & \colhead{($19 - 38\mu$m)} }
\startdata
 Nucleus \& Ring & 2196 s & 2139 s & 6696 s & 930 s \\
 Enuc N &  3294 s & \nodata & 5544 s & 2684 s\\
 Enuc S & 3294 s & \nodata & 6804 s & 2684 s \\
 \enddata
 \label{exposure}
 \end{deluxetable*}

With \textit{Herschel}/PACS \citep{poglitsch10} we are now able to target the most important cooling lines of the warm ISM on physical scales much smaller than previously possible. The  KINGFISH project (Key Insights on Nearby Galaxies: a Far- Infrared Survey with \textit{Herschel} - PI: R. C. Kennicutt) is an open-time \textit{Herschel} key program which aims to measure the heating and cooling of the gaseous and dust components of the ISM in a sample of 61 nearby galaxies with the PACS and SPIRE instruments. The far infrared (FIR) spectral range covered by PACS includes several of the most important cooling lines of the atomic and ionized gas, notably [CII] 158 $\mu$m, [OI] 63 $\mu$m, [OIII] 88 $\mu$m, [NII] 122 $\mu$m, and [NII] 205 $\mu$m. 

NGC~1097 is a Seyfert 1 galaxy with a bright starburst ring with a diameter of 2 kpc and a strong large-scale stellar bar \citep{gerin88,kohno03,hsieh08} with a length of 15 kpc. 
This bar may be responsible for driving gas into the central region of the galaxy, forming a ring of gas near the Inner Lindblad Resonance, triggering the formation of massive star clusters \citep[e.g.][]{athan92}. There is also gas inflow inside the ring possibly fueling the central super-massive black hole \citep{prieto05,fathi06,davies09}, triggering the formation of a compact star cluster seen near the nucleus. The star-forming ring and the nucleus of NGC 1097 are prominent in CO 1--0 \citep{kohno03} as well as ro-vibrational $H_2$ and hydrogen recombination lines \citep{reunanen02, kotilainen00}. Tracers of denser molecular gas such as CO 2--1 and HCN \citep{kohno03} peak at the nucleus and where the dust lanes and the ring intersect. A relatively bright, central concentration of the HCN emission also coincides with the nucleus, caused either by relatively dense molecular gas ($n \sim 10^5$ cm$^{-3}$) or by a strong X-ray radiation field from the nucleus that is affecting the chemistry \citep{meijerink07}. 

Far-infrared photometry of NGC 1097 using PACS was first presented in \citet{sandstrom10}, which showed that the ring dominates the FIR luminosity of the galaxy. Therefore, with PACS it is possible to isolate the ring contribution to FIR luminosity. \citet{sandstrom10} also showed that mid- and far-IR band ratios in the ring vary by less than $\pm20\%$ azimuthally, indicating modest variation in the radiation field heating the dust on ~600 pc scales. \citet{sandstrom10} provides a better estimate of the total bolometric emission arising from the active galactic nucleus and its associated central starburst. The properties of the diffuse gas in the central ring have been investigated by \citet{beirao10} using the PACS spectrograph. They observed an enhancement of [OI] 63 $\mu$m/[CII] 158 $\mu$m ratio and [NII] 122 $\mu$m/[NII] 205 $\mu$m in the ring, which could correspond to an enhancement in the intensity of radiation field and density in the ring. \citet{croxall11} also includes PACS observations of NGC 1097 to study general trends of the ISM heating and cooling.
NGC~1097 is therefore an ideal system to study the physical conditions of the ISM in a nearby galaxy with both a starburst and an active nucleus.
However, the FIR maps used in this study does not probe HII regions, 
nor the warm molecular gas, which can trace shock and AGN excitation, nor do they compare the Herschel data to the emission in the mid-infrared where the strong PAH features reside. These ISM phases are better studied using emission lines found in the mid-infrared ($5 - 40\mu$m). 

In this paper we present a study of the properties of the heating and cooling processes of the ISM in the ring and bar of NGC 1097 based on spectral maps using PACS on board the Herschel Space Observatory and the Infrared Spectrograph (IRS) on board the Spitzer Space Telescope. These two instruments are necessary to cover the wavelength range from $5.5 - 230\mu$m, where the most important tracers of gas heating and cooling are emitted. We will diagnose ISM properties such as PAH ionization and heating efficiency to answer the following questions: how does the gas heating efficiency vary with PAH ionization? Is there a variation of PAH heating efficiency between the ring and the regions at the bar? Is the mechanical energy created by the inflow of gas into the ring an important component of  the gas heating in the ring? In this paper we address these issues in the following sections: in \S2 we present the observations and describe the reduction process of the Spitzer-IRS data into data cubes; in \S3 we present the spectra and the spectral maps of the nucleus and both ends of the bar for several spectral features such as PAHs and $H_2$ lines and use ratio maps to analyze the PAH emission distribution, the gas heating efficiency, and the cause of the warm $H_2$ emission; in \S4 we discuss the implications of the results in \S3. Our main conclusions are summarized in \S5. Throughout the paper we use a distance of 19.1 Mpc for a cosmology $H_0=75$ km/s/Mpc, using the Tully-Fisher relation \citep{willick97}.

\section{Observations and Data Reduction}
\label{obs}

NGC 1097 was a Science Demonstration Phase target observed with the $Herschel-PACS$ instrument as part of the KINGFISH (Key Insights on Nearby Galaxies: a Far Infrared Survey with Herschel; PI: Kennicutt) Open Time program. The galaxy was observed with the PACS integral field unit in both chop-nod and wavelength-switching modes. The observations and data reduction procedure of the PACS spectroscopy data were described in \citet{beirao10}. The observations use an  2 x 2 oversampled raster map strategy (step size $12\arcsec$).  The standard pipeline procedure for this, and part of the HIPE chop-nod pipeline, is to grid these four separate observations onto a common Ra-Dec grid using a smaller pixel scale than the native $9\arcsec.4 \times 9\arcsec.4$ pixels size (in this case $3\arcsec \times 3\arcsec$ sub-pixels). For updates on PACS reduction and a discussion on the differences between observations in chop-nod and wavelength-switching mode, see \citet{croxall11}, which also studies general trends of the ISM heating and cooling.

\begin{figure}
\centering
\plotone{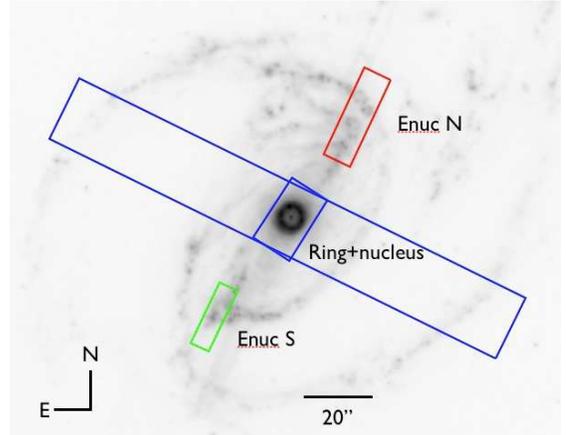}
\caption{Footprints of the Spitzer/IRS spectral maps of NGC 1097, overlaid on a IRAC 8$\mu$m image from the SINGS program (ID 159). The labels refer to the Ring+nucleus (blue), Enuc N (northern edge of the bar, in red) and Enuc S (southern edge of the bar, in green), analyzed in this paper. The Enuc N and Enuc S regions have only short-low observations, while the ring+nucleus have spatial coverage of the IRS with short-low and long-low IRS modules.}
\label{slitsfig}
\end{figure}

NGC 1097 was observed with $Spitzer$-IRS in spectral mapping mode as part of SINGS (Spitzer Infrared Nearby Galaxy Survey; \citet{kenn03}), under program ID 159. The central region which includes the ring and the nucleus was observed using all high- and low-resolution modules, and the edges of the large scale bar, which we refer to as the northern and southern extranuclear regions (Enuc N and S), was observed using all high-resolution modules but with only SL ($5 -14 \mu$m) at low resolution. The central region was observed on July 18, 2004, and the Enuc N and S regions were observed on August 8, 2005. Table 1 summarizes all the observations. In Figure \ref{slitsfig} we show all the footprints of the IRS observations performed on NGC 1097. We extracted only the maps  that contain both modules SL1 and SL2 or LL1 and LL2. While the short-low (SL) module was oriented in a parallel direction relative to the bar, the long-low (LL) module was oriented in a perpendicular direction relative to the bar. The high-resolution modules, SH and LH were oriented at an angle of $45\deg$ relative to the bar. $Spitzer$-IRS spectra of the central region of NGC1097 has been published before by \citet{brandl06} and the \textit{Spitzer}-IRS observations of the nuclear regions for all SINGS galaxies including NGC 1097 are described in \citet{smith07b} (low resolution spectra) and \citet{dale09} (high-resolution spectra). However, these measurements were averaged over large areas. Here we publish the complete IRS spectral maps of NGC~1097 for the first time.
The total exposure times are given in Table \ref{exposure}. 

We used CUBISM \citep{smith07a} to build SL and LL spectral cubes from sets of mapping mode BCD (Basic Calibrated Data) observations taken with Spitzer-IRS. In order to create continuum-subtracted maps of spectral lines and PAH features, we use PAHFIT \citep{smith07b}. PAHFIT is a spectral fitting routine that decomposes IRS low-resolution spectra into broad PAH features, unresolved line emission, and continuum emission from dust grains. PAHFIT allows for the deblending of overlapping features, in particular the PAH emission, silicate absorption and fine-structure lines. This is not possible using CUBISM.  Because it performs a simultaneous fit to the emission features and the underlying continuum, PAHFIT works best if it can fit combined SL and LL spectra across the full $5-40\mu$m range for which it was intended. In order to provide a stable long wavelength continuum from which to estimate the true PAH emission, including the broad feature wings, we formed a combined SL + LL cube by rotating and re-gridding the LL cubes to match the orientation and pixel size of the SL cubes ($1\arcsec.85\times1\arcsec.85$). By calculating the flux in the new LL cubes using a bilinear interpolation (in surface brightness units), we assure that the flux over the native LL pixel scale is conserved. Finally, a small average scale factor of 0.81, calculated by comparing pixels in the spectral overlap region between SL and LL at approximately 14 microns, is applied to the SL data before running PAHFIT on the combined SL + LL cubes. The output of
PAHFIT is saved for each feature, and is used to create the PAH feature maps at the native resolution of the SL data as shown in Fig.~\ref{pahcenfig}.  All
subsequent comparisons of the PAH emission with the PACS data are derived from the SL and LL cubes smoothed to the resolution of the
relevant PACS features, as described in the text, and not from the high-resolution PAH images shown in Fig.~\ref{pahcenfig}. We also used CUBISM to build SH and LH cubes from sets of mapping mode BCD (Basic Calibrated Data) observations and to make maps of [SIII]$18.7\mu$m, [SIII]$33.5\mu$m, and [SiII]$34.5\mu$m.


We also used $Spitzer$-IRAC $8\mu$m and $Spitzer$-MIPS $24\mu$m images of NGC1097, observed as a part of SINGS (for a complete description see \citet{dale05}).The 8 $\mu$m map was corrected for stellar light using a scaled 3.6 $\mu$m image following \citet{helou04}.


\section{Results and analysis}
\label{results}

\subsection{Spectra}
\label{spectrasec}

\begin{figure}
\epsscale{1.2}
\plotone{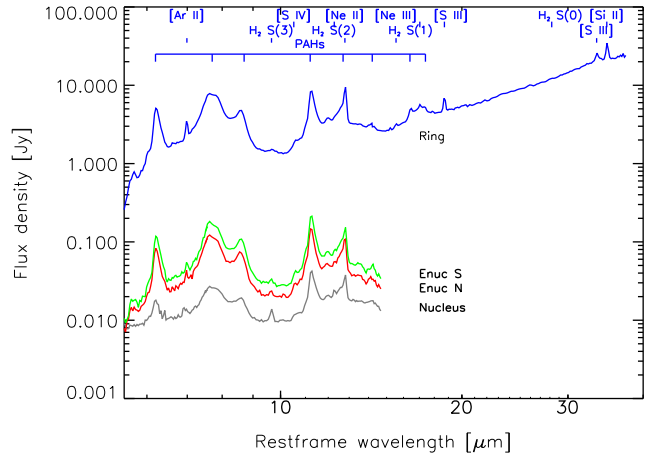}
\caption{Spitzer/IRS low resolution spectra of the ring (blue), Enuc N (red), and Enuc S (green), and the nucleus (grey). All low resolution spectra include the short-low wavelength range (SL; $5-14\mu$m) but only the ring includes IRS long-low (LL; $14 - 38 \mu$m). The Enuc N and Enuc S regions were not observed in LL. The labels indicate the positions of the main spectral features in the mid-infrared.}
\label{regionsfig}
\end{figure}

In Figure \ref{regionsfig} we present the average low-resolution spectra of the four regions highlighted in this study (ring (blue), Enuc N (red), and Enuc S (green)), and a small region centered in the nucleus (in gray). These spectra were produced by extracting a rectangular area from each spectral cube using CUBISM: $17\arcsec \times17\arcsec$ for the ring, and $3.7\arcsec \times3.7\arcsec$ for the nucleus. For Enuc N, the areas are $24.1\arcsec \times40.7\arcsec$ and for Enuc S, the areas are $16\arcsec.7 \times37\arcsec$. The spectrum of the ring is the only one that covers the full $5-38\mu$m range sampled by the SL and LL data. The nuclear spectrum has been subtracted from the ring spectrum in Figure ~\ref{regionsfig} to avoid contamination from the central AGN.
Our low resolution spectra are very similar and show the typical characteristics of star-forming regions: prominent PAH features, a possible absorption feature at $9.7\mu$m and a continuum that rises with wavelength. To make a quantitative comparison between these regions, we used PAHFIT to decompose the low resolution spectra.
We measured the flux of the PAH features, ionic lines and $H_2$ lines, and the results are listed in Tables \ref{pahtab}, and \ref{h2fluxtab}. One of the most prominent differences between the nuclear region and the ring is the flux of the $H_2$ lines at $9.7\mu$m and $12.3\mu$m compared to the PAH features. The flux ratio $H_2$ S(3)/PAH 7.7 $\mu$m is $\sim0.009$ in the nucleus but only $\sim0.004$ in the ring and $\sim0.001$ in the Enuc regions.

\subsection{PAH emission distribution}
\label{pahfigs}

PAHs constitute between $10 - 20\%$ of the interstellar carbon, and were first identified as a possible carrier of the wide emission features at 6.2, 7.7, 8.6, 11.3, and $12.7 \mu$m by \citet{leger84} and \citet{allam85}. These bands are emitted by PAH molecules following photoexcitation. The same PAHs dominate the photoelectric heating of the diffuse ISM. Ionized PAHs are usually associated with the emission bands at $6.2, 7.7, 8.6\mu$m, whereas neutral PAHs are responsible for the emission bands at 11.3, and 12.7$\mu$m. The larger PAHs emit at longer wavelengths \citep[e.g.][]{allam99,hony01}. For this reason the PAH ratios $11.3/7.7\mu$m can be used to diagnose PAH ionization, and $6.2/7.7\mu$m can be used to diagnose grain size \citep[e.g.][]{draine01}, although with caveats, as the $11.3/7.7\mu$m also varies slightly with size.

\begin{figure}
\epsscale{0.8}
\centering
\plotone{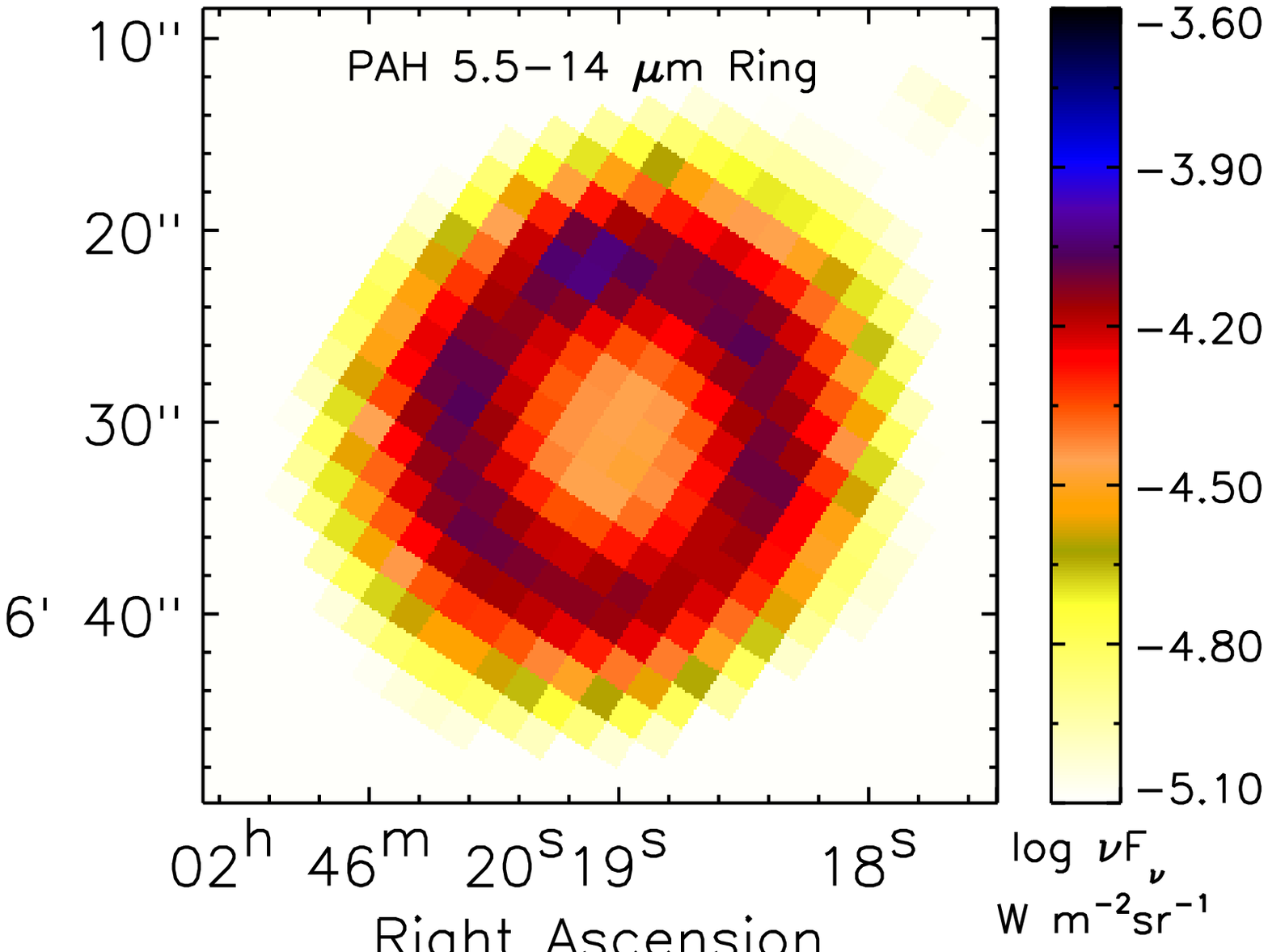}
\plotone{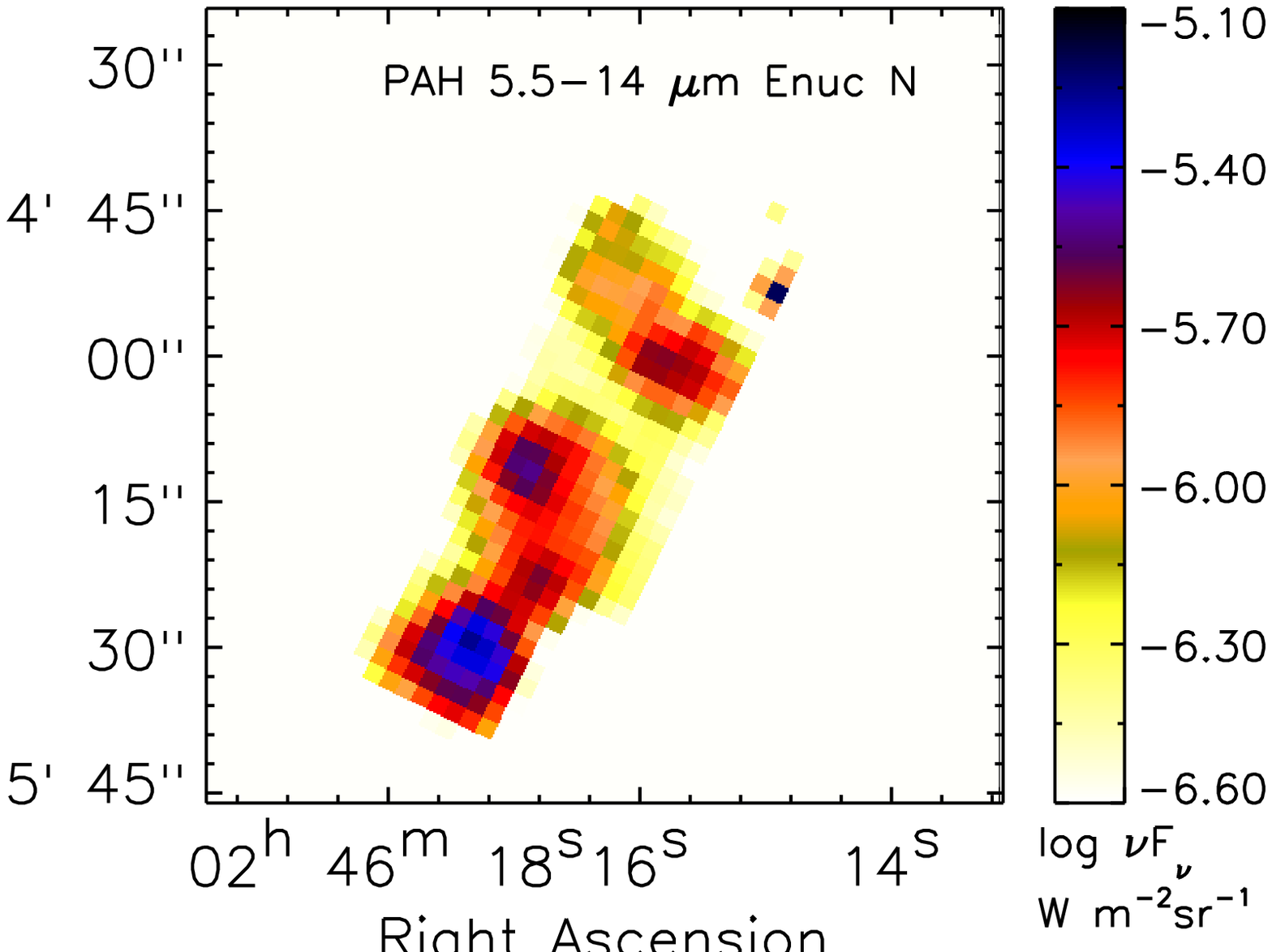}
\plotone{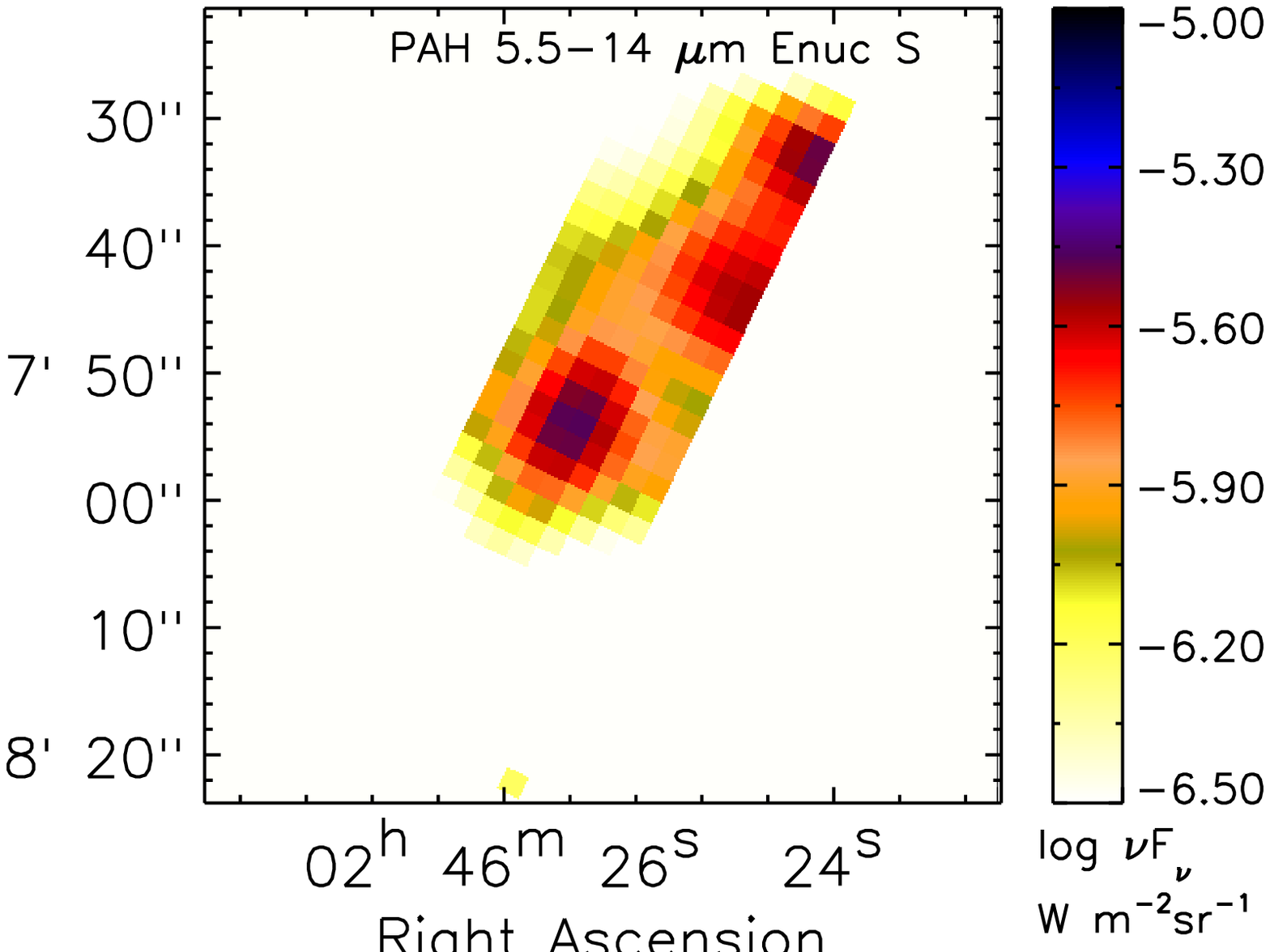}
\caption{Maps of the total PAH emission between $5 -14\mu$m in the ring and nucleus (upper), Enuc N (center), and Enuc S (lower). Each image has a resolution matched to the IRS PSF at $14\mu$m which we approximate as a $4\arcsec.1$ FWHM Gaussian. The pixel scale of $1\arcsec.85$/pixel. North is up.}
\label{pahcenfig}
\end{figure}

In Figure \ref{pahcenfig} we present spectral maps of the continuum-subtracted total PAH emission between $5 -14 \mu$m for the ring, the northern end of the bar (Enuc N) and the southern end of the bar (Enuc S). Each map was constructed by smoothing the 6.2, 7.7, 8.6, 11.3, and $12.7\mu$m to the $14\mu$m resolution, assuming that the PSF at $14\mu$m can be approximated to a gaussian with $FWHM\sim4\arcsec.1$.
The map of the ring reveals that the average PAH surface brightness in the ring is a factor of 10 higher than in the nucleus and outer regions. The PAH surface brightness varies by a factor of $\sim4$ throughout the ring, and peaks in a hotspot north of the nucleus. This is the same hotspot where a maximum of [OI]$63\mu$m/[CII]$158\mu$m was also observed by \citet{beirao10}, which was interpreted as a region with an intense radiation field. PAH flux in the nucleus is a factor of $\sim10$ lower than in the ring, but as seen in Fig. 2 is still quite strong. We can compare this result to other studies which involve PAH observations in the nucleus on NGC 1097 at a higher spatial resolution. \citet{mason07} found no PAH 3.3um in the central $0\arcsec.2$, but they do find strong PAH 11.3um emission over a $3.6\arcsec$ region. As seen in Fig. 2 we also observe strong PAH emission in the nucleus over a similar area, so our results are in agreement.
In the Enuc N and S, the PAH surface brightness varies by a factor of $\sim10$, peaking near star forming regions.

\begin{figure}
\epsscale{0.8}
\plotone{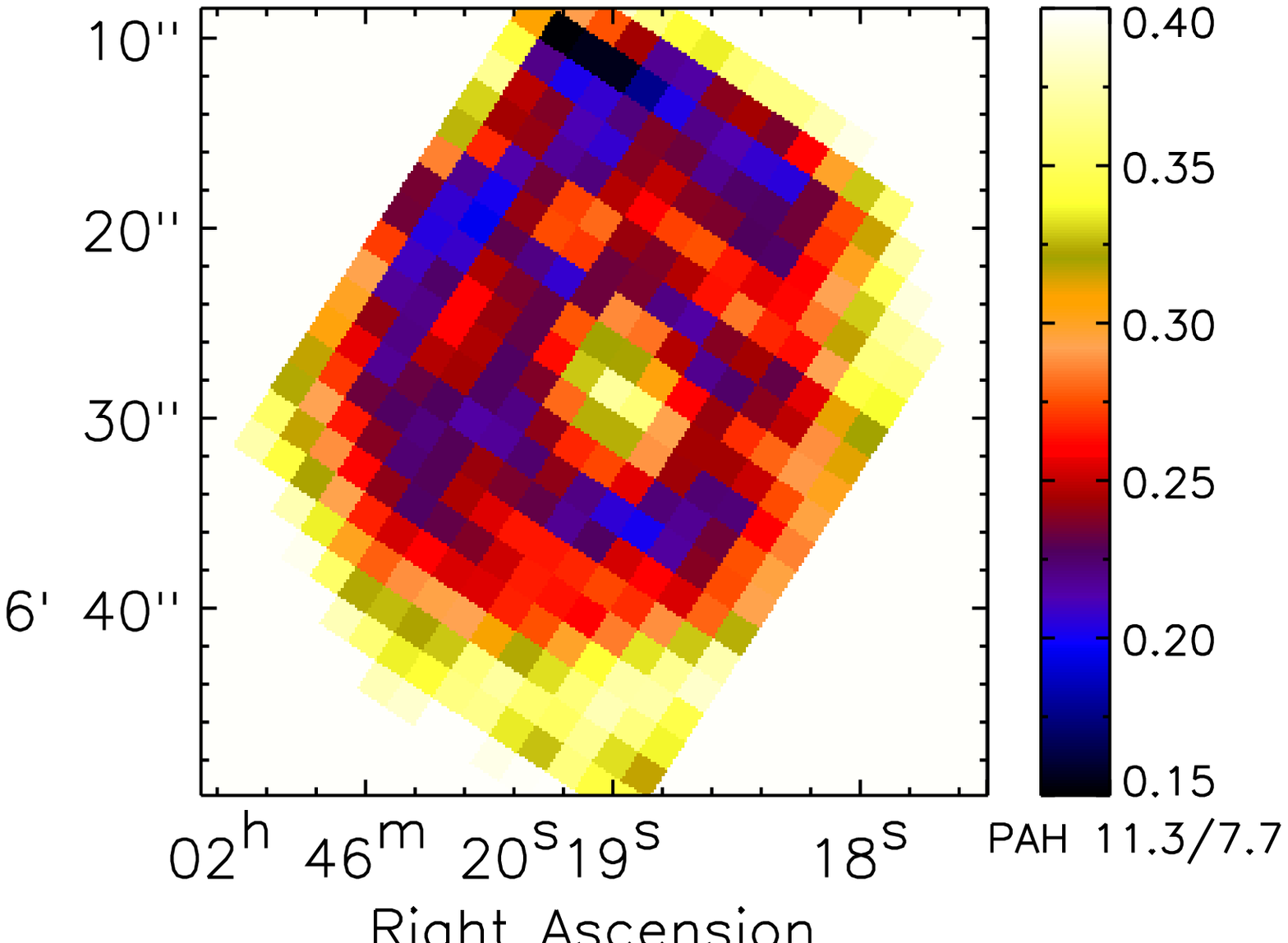}
\plotone{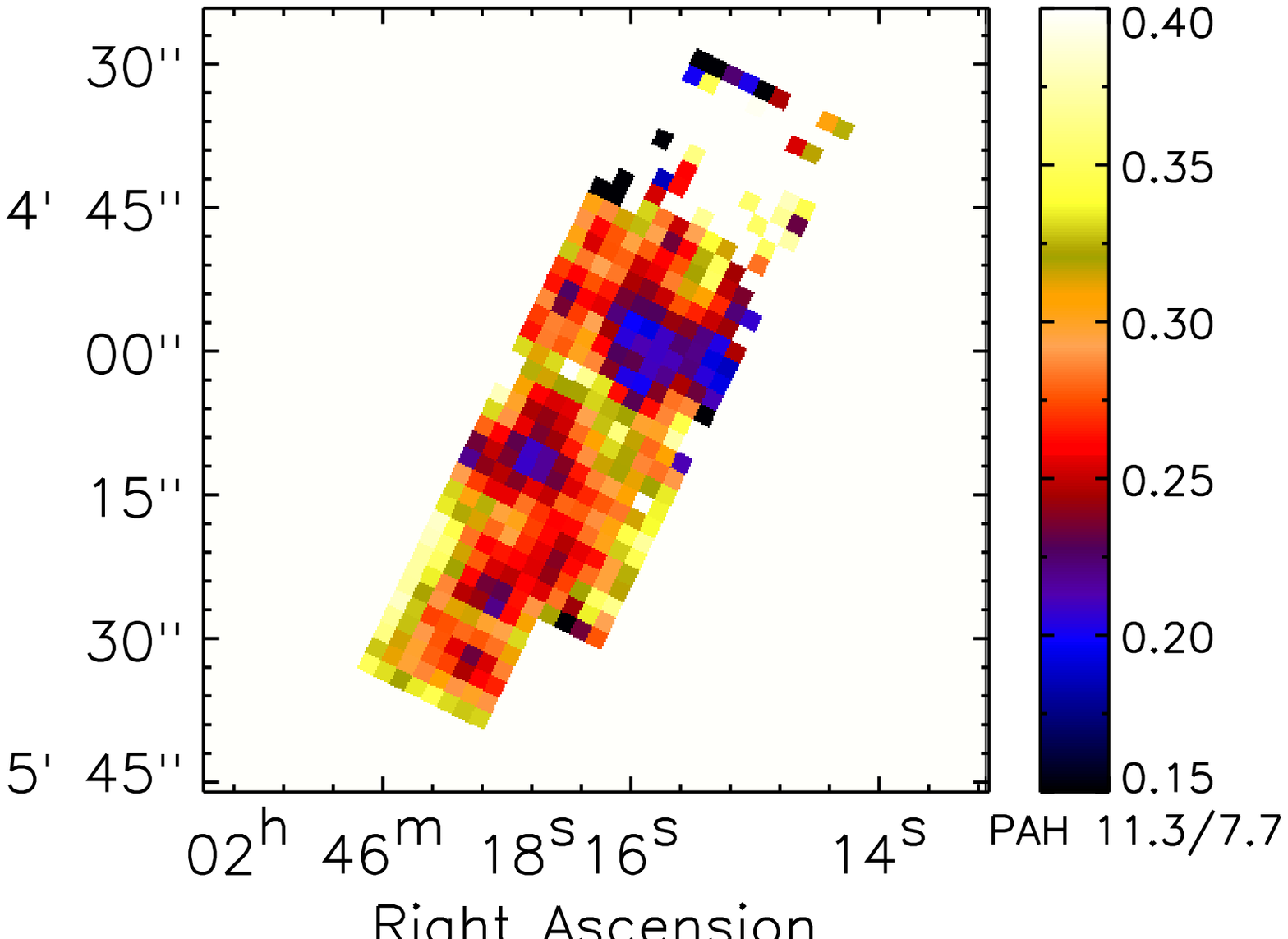}
\plotone{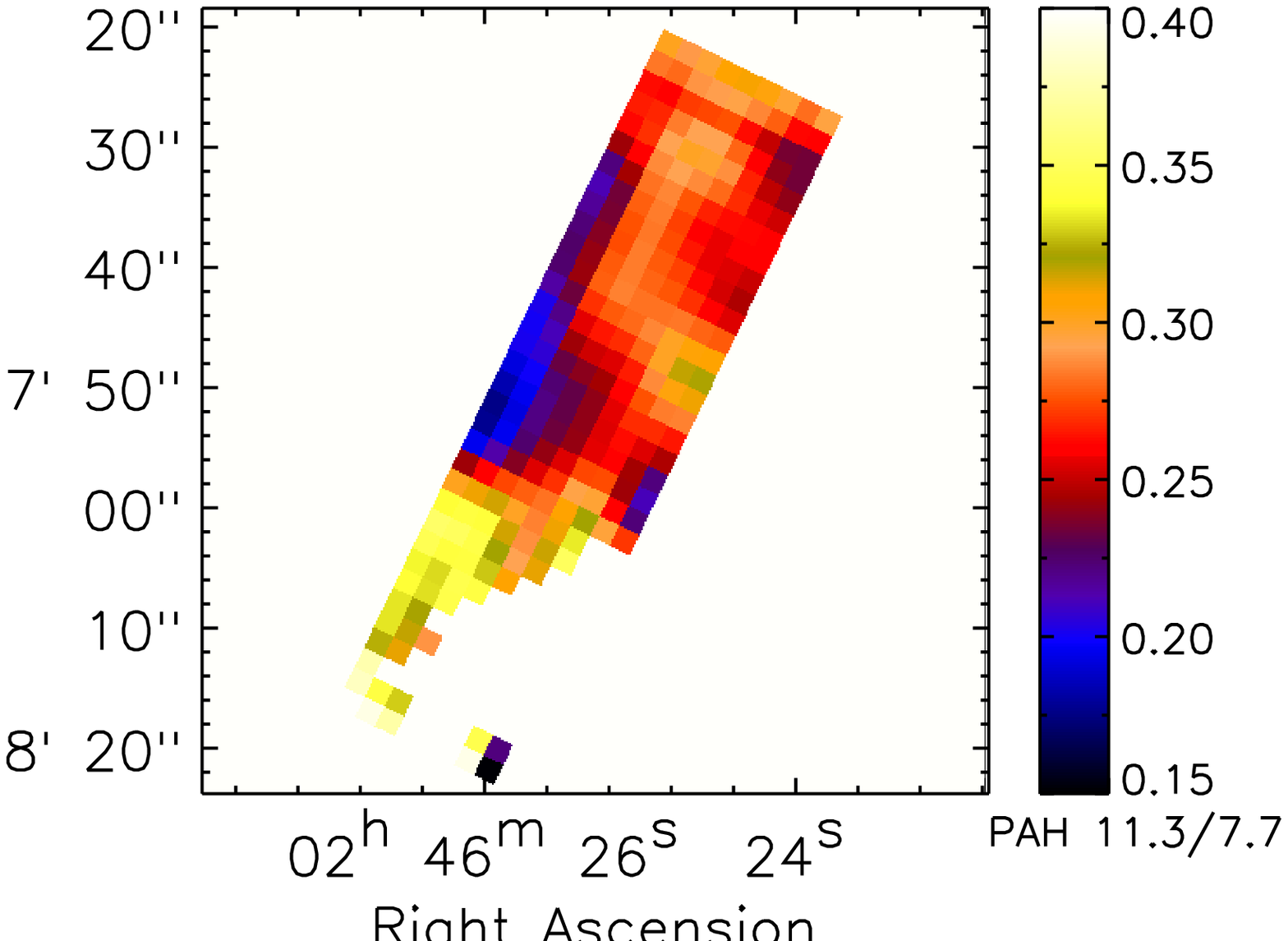}
\caption{Emission maps of the $11.3/7.7\mu$m PAH ratio in the ring and nucleus (upper), Enuc N (center), and Enuc S (lower). Each image has a resolution matched to the IRS PSF at $14\mu$m which we approximate as a $4\arcsec.1$ FWHM Gaussian. The pixel scale of $1\arcsec.85$/pixel. North is up.}
\label{pahratfig}
\end{figure}

\begin{figure}
\epsscale{1.2}
\plotone{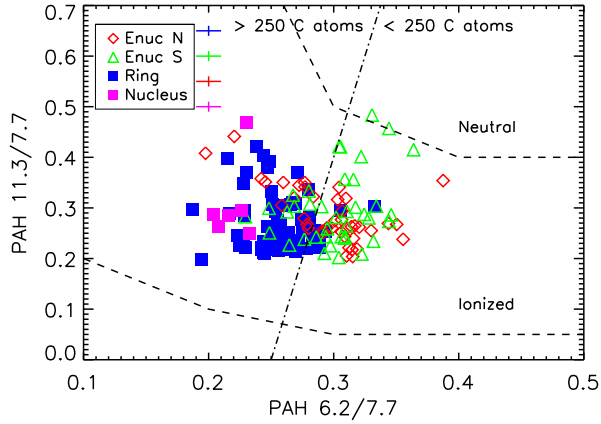}
\caption{Plot of the $11.3\mu$m/$7.7\mu$m PAH flux ratio as a function of $6.2\mu$m/$7.7\mu$m PAH flux ratio. Each point corresponds to a $3\arcsec.7\times3\arcsec.7$ pixel region in the SL maps. The dashed lines indicate ionization states of the PAH molecules. Data from the ring is in blue, the nucleus is in violet, Enuc N is in red, and Enuc S is in green. The error bars indicate the average uncertainties for each region. The dash-dotted line indicates the position of the molecules formed by 250 carbon atoms.  This figure is adapted from \citet{draine01}.}
\label{drainefig}
\end{figure}

In Figure~\ref{pahratfig} we show spectral maps of the $11.3/7.7\mu$m PAH ratios of the ring and extranuclear regions at both ends of the bar. These maps were produced using $7.7\mu$m and $11.3\mu$m PAH maps smoothed to a $4\arcsec.1$ resolution using a gaussian kernel. The ratio increases up to a factor of 2 from a low of $\sim0.2$ in the ring to a peak of $\sim0.4$ outside of the ring and in the nucleus. In the extranuclear regions the ratio varies in the same proportion, with the lowest ratio coinciding with the star forming regions. 
However, in Enuc N the PAH surface brightness peak at the southern end of the map does not coincide spatially with a low value of $11.3/7.7\mu$m PAH ratio. This is surprising, but the reason is not clear, and indicates that the PAH ratios are not always a direct indicator of the starlight intensity. 
In Figure \ref{drainefig} we plot the $11.3/7.7\mu$m PAH ratio as a function of $6.2/7.7\mu$m using PAH 6.2, 7.7, and 11.3 maps for the ring (blue), Enuc N (red), and Enuc S (green). These maps were convolved to the resolution at $11.3\mu$m, $2\arcsec.7$, assuming a gaussian PSF. Each point corresponds to a $3\arcsec.7\times3\arcsec.7$ region. The tracks indicate $6.2/7.7\mu$m and $11.3/7.7\mu$m PAH ratios for neutral and ionized PAH grains \citep{draine01}, and the dot-dashed line indicates where PAH grains have approximately 250 C atoms.  The error bars indicate representative measurement uncertainties. The $11.3/7.7\mu$m values for the ring are concentrated at $\sim0.23$, but there is a wider dispersion for the Enucs. The $6.2/7.7\mu$m ratio is on average 20\% larger in the Enuc N and S than in the ring, but 20\% lower in the nucleus than in the ring. This suggests that on average PAH grains may be smaller in the Enuc regions, but larger in the nucleus than in the ring. Table 2 shows the integrated fluxes of the main PAH features and PAH ratios in the ring, nucleus, Enuc N and Enuc S.

\subsection{Fine-Structure Lines and PAH Emission}
\label{heating}


\begin{figure}
\plotone{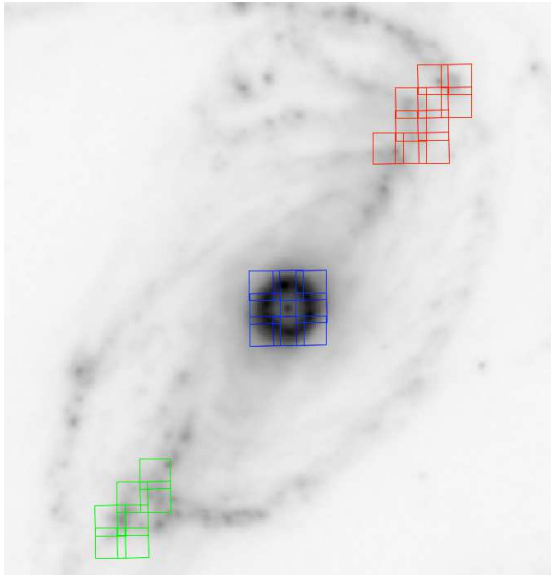}
\caption{The $12\arcsec\times12\arcsec$ square apertures selected for measuring the [CII]$158\mu$m, [OI]$63\mu$m, PAH ($5.5 - 14\mu$m), $70\mu$m, and $100\mu$m fluxes in the ring (blue), Enuc N (red), and Enuc S (green), overlaid on a IRAC $8\mu$m image of NGC 1097. North is up.}
\label{regfig}
\end{figure}

\begin{figure}
\epsscale{2.3}
\plottwo{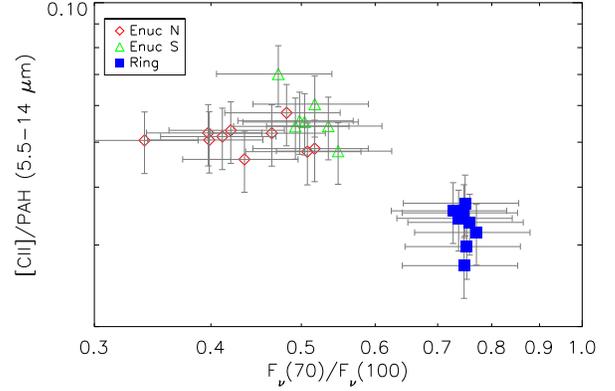}{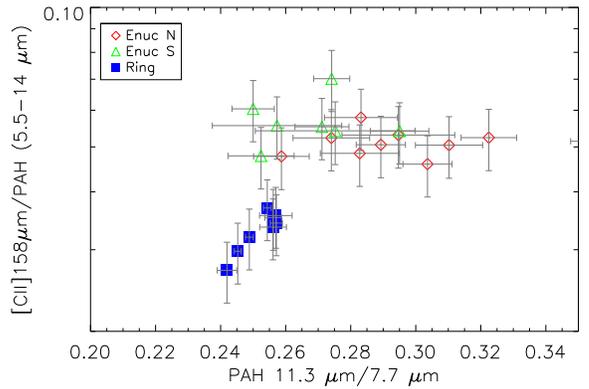}
\caption{Top: Plot of the ratio [CII]$158\mu$m/PAH ($5.5 - 14 \mu$m) as a function of the $70\mu$m/$100\mu$m flux ratio. 
Bottom: Plot of the ratio [CII]$158\mu$m/PAH ($5.5-14 \mu$m) as a function of the $11.3/7.7$ PAH ratio. Each point was calculated for a resolution of $11\arcsec$ per beam.
}
\label{ciipahfig}
\end{figure}

\begin{figure}
\epsscale{1.2}
\plotone{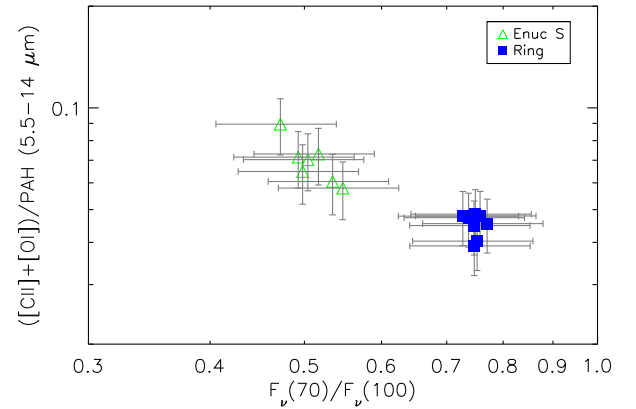}
\caption{Plot of the ratio ([CII]$158\mu$m+[OI]$63\mu$m)/PAH ($5.5 - 14 \mu$m) as a function of the $70\mu$m/$100\mu$m flux ratio. Each point was calculated for a resolution of $11\arcsec$ per beam.}
\label{oipahfig}
\end{figure}

For a direct comparison between the PACS 
and the IRS data, we measured the flux of the [CII]$158\mu$m and [OI]$63\mu$m fine-structure lines, the total flux of the PAH emission features between $5.5 - 14\mu$m, the $11.3/7.7\mu$m PAH ratio, and the $70\mu$m/$100\mu$m ratio in the ring and extranuclear positions over a grid composed of $12\arcsec\times12\arcsec$ square regions, shown in Figure \ref{regfig}. We chose these regions to properly sample the flux at the longest wavelength where the FWHM is close to $11\arcsec$.
Before measuring the total PAH flux and the $11.3/7.7\mu$m PAH ratio, we made convolution kernels to smooth each PAH feature map at $6.2, 7.7, 8.6, 11.3, 12.6$, and $14.1\mu$m to the $Herschel$ beam at $160\mu$m, using PSFs calculated by \citet{aniano11}, and theoretical PSFs for IRS made with STinyTim for each wavelength above. We then rebinned all maps to match the pixel size used in the PACS FIR spectral maps ($3\arcsec$/pixel). The PAH 7.7$\mu$m and the $11.3\mu$m fluxes were measured summing all pixels within each region in Fig.~\ref{regfig}. We also used [OI]$63\mu$m maps and the $70\mu$m and $100\mu$m PACS images convolved to the PACS $160\mu$m PSF using the convolution recipe described in \citet{aniano11}.
Due to low S/N, [OI]$63\mu$m could only be measured in the ring and Enuc S. The total luminosity of [CII]$158\mu$m and [OI]$63\mu$m in the ring is $L_{[CII]}\sim1.4\times10^8 L_{\odot}$ and $L_{[OI]}\sim5.4\times10^7 L_{\odot}$, respectively. In Table~\ref{regiontab} we present all the [CII], [OI], and PAH fluxes measured for all the regions, as well as ionic lines measured at hi-res, [SIII]$18\mu$m, [SIII]$33.5\mu$m, and [SiII]$34.5\mu$m.

In Figure \ref{ciipahfig} (upper) we plot the ratio [CII]$158\mu$m/PAH (5.5 -14$\mu$m) as a function of $70\mu$m/$100\mu$m flux ratios, based on PACS photometry for the ring and the two extranuclear maps at a resolution of $11\arcsec$ per beam. The $70\mu$m/$100\mu$m flux ratio is a commonly used indicator of dust temperature. Each datapoint is one of the $12\arcsec$ square regions ($\sim1.1$ kpc). There is a clear difference between the ring and the extranuclear positions in both [CII]$158\mu$m/PAH (5.5 -14$\mu$m) and $70\mu$m/$100\mu$m. We performed a KS test to determine the significance of this difference. The D statistic is 0.85, with a probability of 0.029\% of the ring and Enuc S belonging to the same population. The difference between the two regions is therefore significant. For the ring, the median [CII]$158\mu$m/PAH (5.5 -14$\mu$m) is $\sim0.034$ and for the Enucs $\sim0.053$, about a factor of $\sim1.6$ above the median ring value. The $70\mu$m/$100\mu$m ratio ranges between 0.35-0.55 for Enuc N and S and between 0.70-0.80 for the ring.
The dust is warmer and the [CII]$158\mu$m/PAH (5.5 -14$\mu$m) is lower in the ring than in the Enucs. 
If we consider the total infrared emission between $5 - 10\mu$m, F (5.5 -10$\mu$m), we can compare our results with \citet{helou01}, in which [CII]$158\mu$m/F (5.5 -10$\mu$m) is constant, suggesting that photoelectrons liberated from PAH grains were responsible for heating the gas. Our values for [CII]/F($5-10\mu$m) are $\sim50\%$ lower in the ring than in the Enucs.
In Figure \ref{ciipahfig} (lower) we plot the same ratio [CII]$158\mu$m/PAH (5.5 -14$\mu$m) now as a function of the $11.3/7.7\mu$m PAH ratio, at a common resolution of $11\arcsec$ per beam. The distribution of $11.3/7.7\mu$m PAH ratios in the ring is narrower than in the Enuc regions, and has a median value of 0.25, lower than the median value of $11.3/7.7\mu$m in the Enuc regions, which is 0.28. This could be due to the fact that the PAH emission in the ring is more ionized. We performed a Spearman rank correlation test, and the result is 0.33, revealing a weak correlation between [CII]$158\mu$m/PAH (5.5 -14$\mu$m) flux ratio and the $11.3/7.7\mu$m PAH ratio. However, this apparent correlation is only due to the lower [CII]$158\mu$m/PAH (5.5 -14$\mu$m) values in the ring, whereas the Enuc values of [CII]158$\mu$m/PAH (5.5-14$\mu$m) appear to be insensitive to the local 11.3/7.7$\mu$m flux ratio. Therefore it is not clear that the difference in [CII]$158\mu$m/PAH (5.5 -14$\mu$m) between the ring and the Enucs is due to a variation of the PAH ionization. Note that the $11.3/7.7\mu$m datapoints shown in Figure~\ref{ciipahfig} were measured over a much wider area, and therefore are an average of the datapoints in Figures 4 and 5. This accounts for the narrower range of $11.3/7.7\mu$m values in Figure~\ref{ciipahfig}.
[OI]$63\mu$m can also be an important coolant in warmer regions. In Figure~\ref{oipahfig} we plot the sum of the [CII]$158\mu$m and [OI]$63\mu$m emission over the PAH flux between $5-14\mu$m against the $70\mu$m/$100\mu$m flux ratio. We measured the [OI]$63\mu$m fluxes after smoothing the [OI]$63\mu$m image to the 160$\mu$m PACS PSF (a resolution of approximately $11\arcsec$), using kernels from \citet{aniano11}. Here we exclude Enuc N, due to low S/N in the [OI]$63\mu$m line. The gap between the Enuc S and the ring in ([CII]$158\mu$m+[OI]$63\mu$m)/PAH is smaller than in [CII]$158\mu$m/PAH, with a median of $\sim0.047$ for the ring and $\sim0.070$ for Enuc S. [OI]$63\mu$m has therefore a much larger role in the cooling of the gas in the ring than in Enuc S. This is consistent with the behavior of ([CII]$158\mu$m+[OI]$63\mu$m)/PAH in the galaxies as reported by \citet{helou01}. Both the ring and the Enucs in NGC 1097 are within the range of parameters studied in \citet{helou01}. They calculate [CII]/F($5-10\mu$m) for a sample of star-forming galaxies using global averages, and shows no trend with $60\mu$m/$100\mu$m. However, they find a large scatter in ([CII]+[OI])/F($5-10\mu$m) as a function of $60\mu$m/$100\mu$m. What we are uncovering is the inherent variations within galaxies that might drive the scatter in this relation. We should note that all the measurements are affected by the uncertainty on the PACS beam size and the shape of the PSFs, quantified in \citet{aniano11}. 

Another important cooling line is [SiII] at $34.8\mu$m. With a critical density $\sim10^5$ cm$^{-3}$ for collisions with neutral hydrogen and $\sim10^2$ cm$^{-3}$ for collisions with electrons, and a low ionization potential of 8.15 eV, this line comes from a wide variety of regions, such as ionized gas and warm atomic gas, photodissociation regions and X-ray dominated regions \citep {hollen99}. HII regions represent a large fraction of [SiII]$34.8\mu$m emission in the ring (between 30\%-60\%, \citet[e.g.][]{nagao11}). We used an [SiII]$34.8\mu$m image of the ring and Enuc S, and convolved it to the $160\mu$m Herschel-PACS beam using theoretical $Spitzer-IRS$ PSFs made with STinyTim and $Herschel$ PSFs calculated by \citet{aniano11}. From these images we measured a luminosity $L_{[SiII]}\sim4\times10^8L_{\odot}$ in the ring and $L_{[SiII]}\sim4\times10^7L_{\odot}$ in Enuc S, so $L_[SiII]\sim1.6\times L_{[OI]}$. This extra cooling would raise the ring and Enuc S in Figure \ref{oipahfig}.



\begin{figure}
\epsscale{2.3}
\plottwo{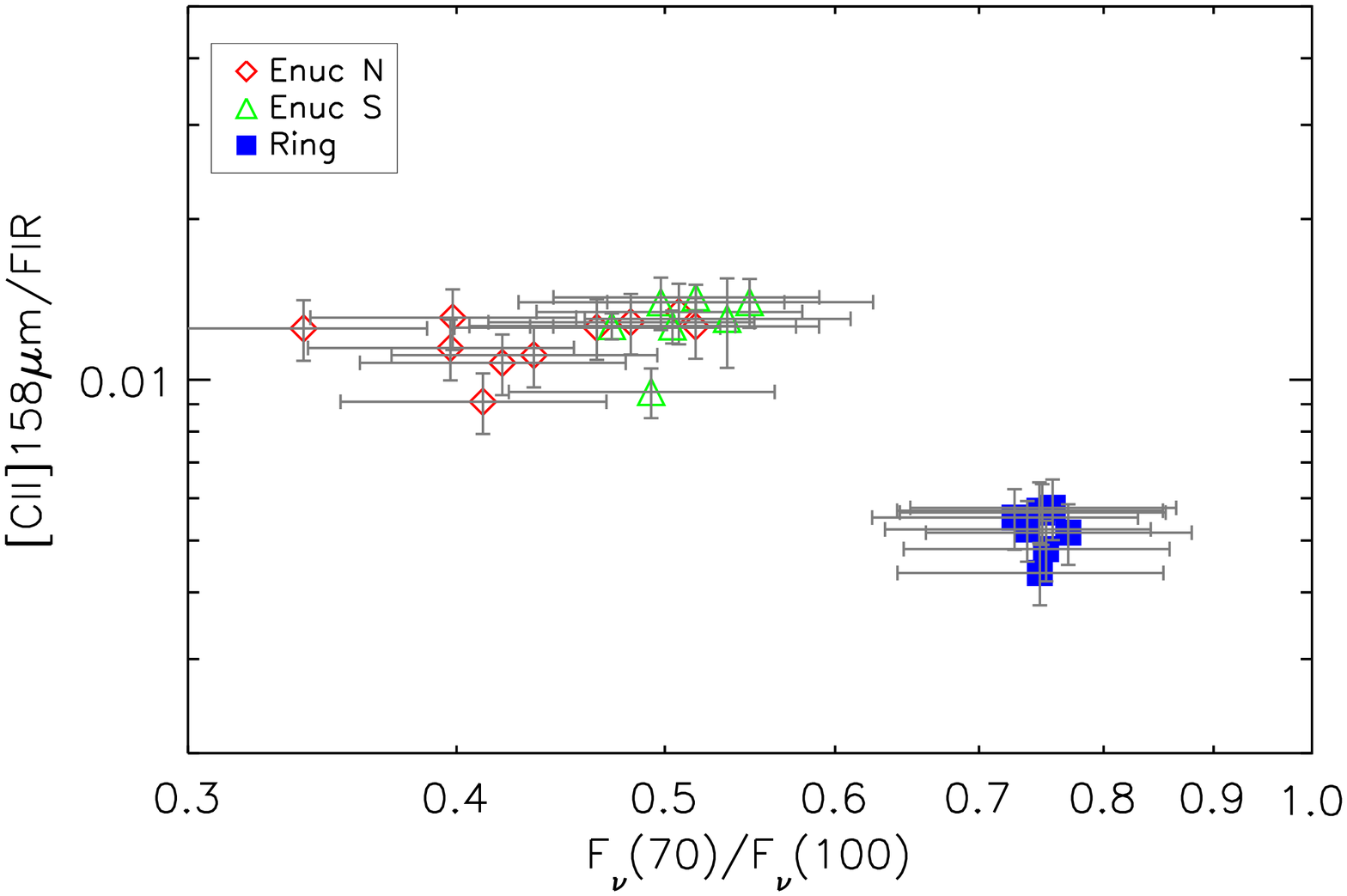}{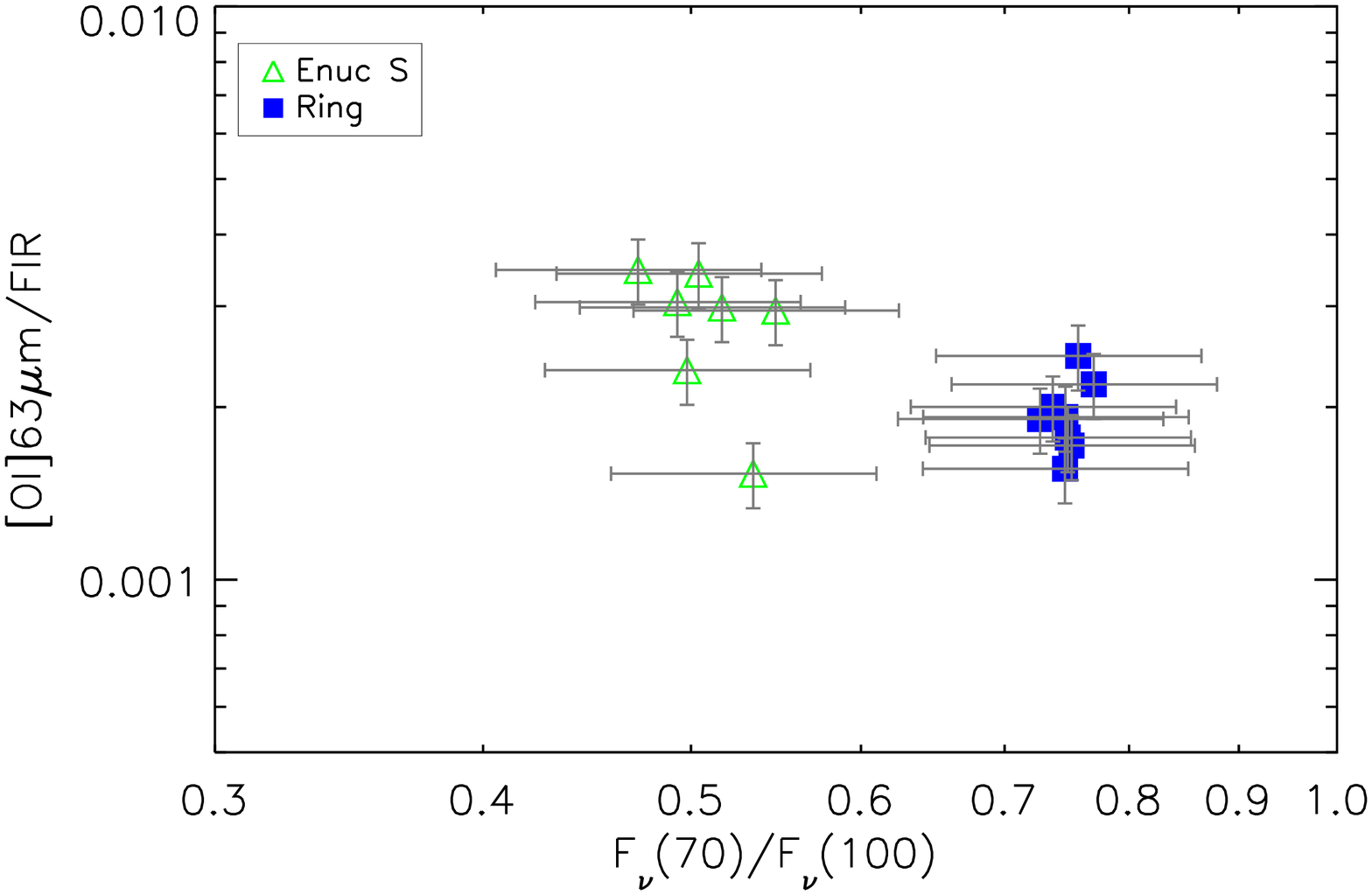}
\caption{Top: Plot of [CII]$158\mu$m/FIR as a function of the $70\mu$m/$100\mu$m flux ratio. Bottom: Plot of [OI]$63\mu$m/FIR as a function of the $70\mu$m/$100\mu$m flux ratio. }
\label{ciifirfig}
\end{figure}

In Figure~\ref{ciifirfig} we show a plot of [CII]$158\mu$m/FIR as a function of the $70\mu$m/$100\mu$m flux ratio, with the datapoints calculated at a common beam size ($11\arcsec$). FIR is the far infrared luminosity, estimated as FIR=TIR/2, with TIR as the total infrared luminosity, which was calculated using the formula by \citet{dale02}, replacing the 70$\mu$m, and 160$\mu$m flux by PACS $70\mu$m and $160\mu$m fluxes. The ring has a [CII]$158\mu$m/FIR which is a factor of $\sim3$ lower than the Enuc regions, and this difference is much larger than the factor of $\sim1.5$ in [CII]$158\mu$m/PAH (5.5 -14$\mu$m) between the ring and the Enucs. This effect has been associated with regions with high $G_0/n$ where other cooling lines are enhanced, such as [OI]$63\mu$m \citep{malhotra01}. The [CII]$158\mu$m/FIR in the ring is similar to the value in other star-forming galaxies observed by ISO such as NGC 1482, which has [CII]$158\mu$m/FIR$\sim0.004$ and NGC1569, with [CII]$158\mu$m/FIR$\sim0.0027$ \citep{malhotra01}. In Figure \ref{ciifirfig} (bottom) we plot [OI]$63\mu$m/FIR for the ring and Enuc S. We see that [OI]$63\mu$m/FIR is still lower for the ring, but [OI]$63\mu$m contributes at most $\sim25\%$ of the [CII]$158\mu$m+[OI]$63\mu$m flux in the Enucs and $\sim30\%$ of the [CII]$158\mu$m+[OI]$63\mu$m flux in the ring. Although some of the cooling indeed goes through [OI]$63\mu$m, it cannot fully compensate for the difference between the ring and the Enuc S in [CII]$158\mu$m/FIR. The decrease in [CII]$158\mu$m/FIR can be caused not only by cooling via other emission lines, but also by absorption of ionizing photons by dust, low PAH abundance due to PAH destruction or coagulation into larger grains, collisional suppression of [CII]$158\mu$m, or the [CII]$158\mu$m being optically thick.

\subsection{Warm $H_2$ emission}
\label{h2}

\begin{figure*}
\plotone{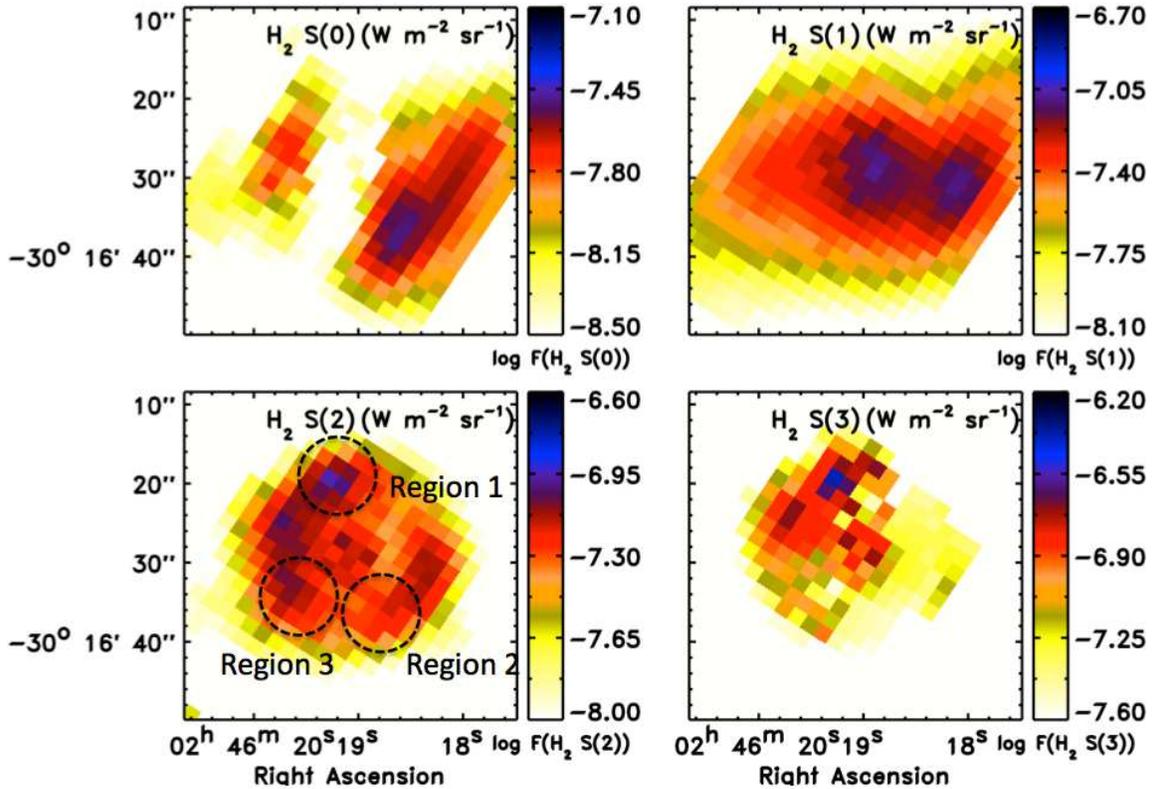}
\caption{Map of the $H_2$ S (0) and $H_2$ S(1), $H_2$ S(2), and $H_2$ S(3) at $28.2\mu$m, $17.0\mu$m, $12.3\mu$m, and $9.7\mu$m respectively.  ``Region 1", ``Region 2", and ``Region 3" denote three regions selected along the ring (see description in text). All images are at their native resolutions, i.e. $6\arcsec.8$, $4\arcsec.1$, $3\arcsec.1$, and $2\arcsec.4$, respectively if we approximate the IRS PSFs as Gaussians. North is up.}
\label{h2fig}
\end{figure*}

Figure \ref{h2fig} shows flux maps of the $H_2$ rotational lines from S(0) to S(3) at their native resolution, extracted pixel-by-pixel with PAHFIT. Most of the S(0) flux is emitted from two spots in the ring that coincide with the intersection of the bar with the ring. The S(1) line flux is also emitted from these spots, but also peaks at the nucleus. In S(2) and S(3) the flux peaks at a region in the NE part of the ring, where the main H$\alpha$ and FIR lines are emitted \citep{beirao10}, but there is also some emission from the nucleus. 


\begin{figure}
\epsscale{2.4}
\plottwo{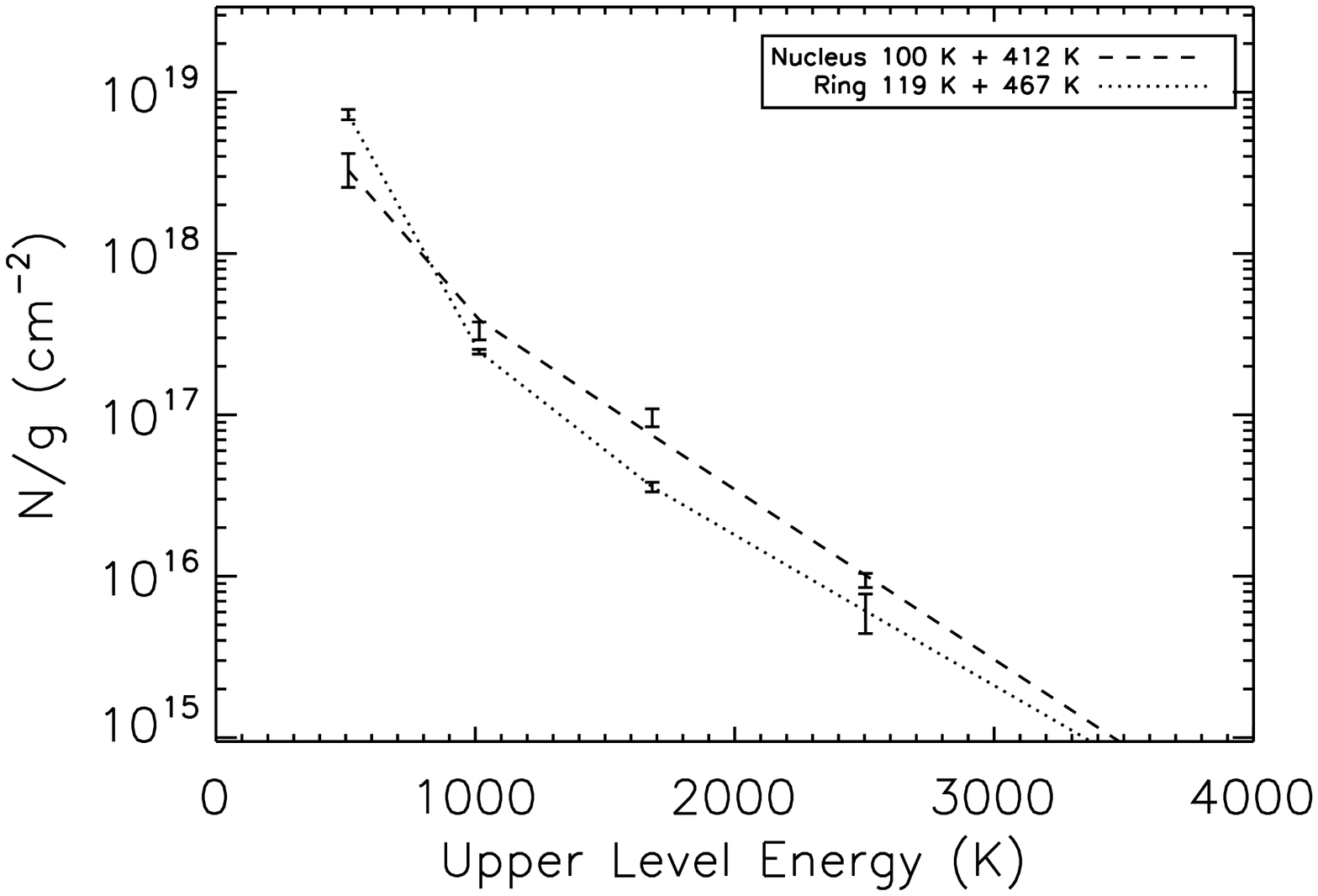}{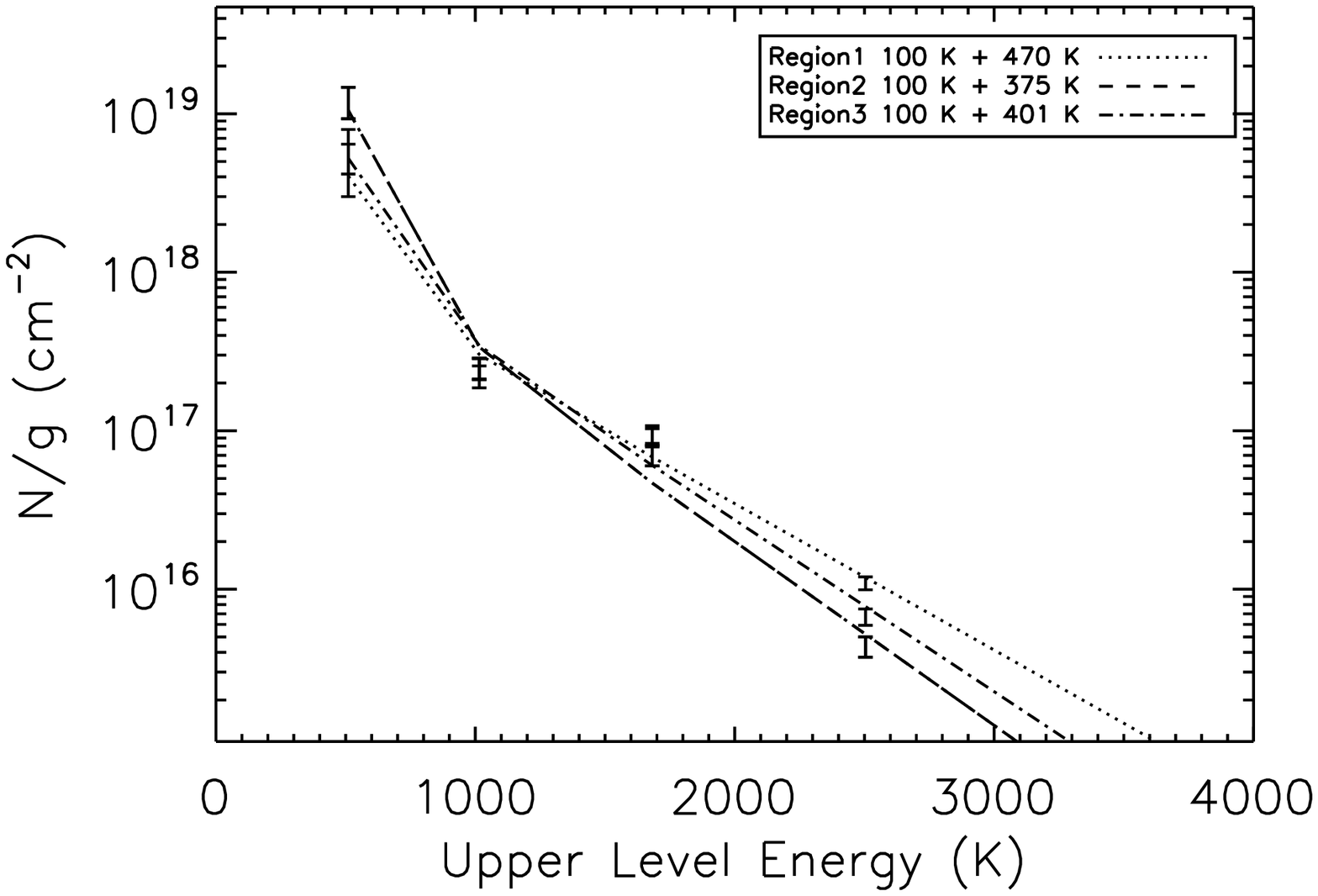}
\caption{Top: mid-infrared $H_2$ excitation diagrams for the nucleus and the ring of NGC 1097. Bottom: excitation diagrams for the three regions selected along the ring and highlighted in Figure \ref{h2fig}. The lines indicate the best fit to the $H_2$ datapoints. (see Table \ref{h2tab}). }
\label{excitfig}
\end{figure}

With the $H_2$ emission maps, we can estimate the gas temperature, column density, and mass distributions by fitting the $H_2$ fluxes using the Boltzmann equation \citep{rigo02}, assuming that the ortho/para ratio is in local thermodynamical equilibrium (LTE).
In Figure \ref{excitfig} we plot the excitation diagrams for the ring and nucleus of NGC 1097 (upper), and three selected ($\sim10\arcsec$) regions on the ring (lower), with fits to the warm and hot $H_2$ phases. 
In this fit, 100 K was set as the lower temperature limit, as this is approximately the lowest temperature in a collisionally excited $H_2$ gas. Before the fit we convolved the S(3), S(2) and S(1) images to a gaussian with $FWHM=7\arcsec$, which is the resolution at $28\mu$m. In Table \ref{h2tab} we present the properties of the warm $H_2$ gas in the ring, nucleus, and three $10\arcsec$ diameter regions in the ring indicated in Figure \ref{h2fig}. The lower temperature component in the nucleus has $\sim2\%$ of the total $H_2$ luminosity of the nucleus, and therefore almost all the $H_2$ lines can be fitted with a single temperature. The ring is fitted with two components, a cool component with a temperature of 119$\pm$11 K and a hot component with 467$\pm79$ K. In the ring, the cool component emits $20\%$ of the $H_2$ flux, so there is a large quantity of gas in the ring at $\sim100$ K that is not present in the nucleus.
The lower plot in Figure \ref{excitfig} shows that the hot component varies between $T\sim375$ and 470 K. Of the three selected regions, Region 2 has the highest fraction of emission from the cool component, $\sim16\%$, and the lowest temperature of the warm component, perhaps indicative of a reservoir of colder gas in this region.  

\section{Discussion}


\subsection{Physical conditions in the gas}
\label{pdr}

To explore the physical conditions of the gas in the ring and Enuc regions we use \citet{kaufman06} PDR models to determine the intensity of the radiation field and the density of the neutral gas using the [OI]$63\mu$m and [CII]$158\mu$m fine structure lines. Our measurements of [CII]$158\mu$m include an ionized gas contribution, where the gas is emitted from diffuse ionized regions. Therefore one needs to subtract the ionized contribution to [CII]$158\mu$m.
This contribution can be estimated from measurements of the [NII] $122\mu$m and [NII]$205\mu$m lines, which arise in diffuse ionized gas and HII regions. 
The [NII]$122\mu$m/[NII]$205\mu$m values published by \citet{beirao10} were affected by an incorrect Relative Spectral Response Function (RSRF) estimation, which affects the calibration of the continuum at $\lambda>200\mu$m. To correct this effect we have to multiply the [NII]$205\mu$m flux by a correction factor of 4.5, which results in a [NII]$205\mu$m/$122\mu$m$\sim0.25$ in the ring.

Using the estimated [NII]$205\mu$m/[NII]$122\mu$m value for the ring, we first calculate the fraction of [NII]$122\mu$m from diffuse ionized gas ($n_e=10$ cm$^{-3}$). \citet{rubin94} calculate [NII]$205\mu$m/[NII]$122\mu$m$\sim 1$ for $n_e=10$ cm$^{-3}$ and [NII]$205\mu$m/[NII]$122\mu$m $\sim 0.1$ for $n_e=10^3$ cm$^{-3}$. The average value of the [NII]$205\mu$m/$122\mu$m in the ring is $\sim0.25$. Therefore we estimate that the diffuse medium contributes $\sim17\%$ of the observed [NII]$122\mu$m emission in the ring.
Using C$^+$ and N$^+$ abundances from \citet{meyer97, sofia04}, $A$ values from \citet{galavis97, galavis98} and collision strengths from \citet{hudson05, blum92}, we derived a [CII]158$\mu$m/[NII]122$\mu$m$\sim4.8$ in diffuse ionized gas and $[CII]158/[NII]122\sim0.8$ in the dense ionized gas. If 17\% of the observed [NII]122$\mu$m is emitted from the diffuse ionized gas, [CII]$158\mu$m$_{ion}\sim1.5 $[NII]$122\mu$m. We can use this estimate to calculate the ionized gas contribution to [CII]$158\mu$m. From \citet{beirao10}, [CII]$158\mu$m/[NII]122$\sim4.5$ in the ring, which suggests that $\sim33\%$ of the [CII]$158\mu$m flux in the ring is emitted from ionized gas and $\sim67\%$ from PDRs. Note that this estimate is affected by the uncertainty on the [NII]$205\mu$m/$122\mu$m value, which is $\sim50\%$.

Similarly, we can determine the maximum ionized gas fraction of the [CII]$158\mu$m emission in Enuc S, and compare to the value for the ring estimated above. Because of low S/N, no measurement of [NII]$205\mu$m was possible in Enuc N and S and no [NII]$122\mu$m was possible in Enuc N.
However, we can use [SIII]$18\mu$m/$33\mu$m ratio as a tracer of density for regions with $n_e>100$ cm$^{-3}$, and use the density estimated from this ratio to estimate the corresponding [NII]$122\mu$m/[NII]$205\mu$m \citep[see][]{croxall11}. We measure [SIII]18/33$\sim1.2$ in Enuc S. This corresponds to [NII]$122\mu$m/[NII]$205\mu$m$\sim8$. Assuming 10 cm$^{-3}$ for the diffuse component and 10$^3$ cm$^{-3}$ for the dense component, we can estimate the diffuse gas contribution to [NII]$122\mu$m/[NII]$205\mu$m in Enuc S as $\sim2.8\%$. 
Using these values, we get [CII]$158\mu$m/[NII]122$\sim0.9$ for the ionized component in Enuc S, which is $\sim5\%$ of the observed ratio, [CII]$158\mu$m/[NII]122$\sim17$ \citep{beirao10}. Therefore, the ionized gas component accounts for $\sim5\%$ of the [CII]$158\mu$m flux in Enuc S. Enuc S therefore seems to have a much lower fraction of the observed [CII]$158\mu$m emission coming from ionized regions than does the ring. 

\begin{figure}
\epsscale{1.2}
\plotone{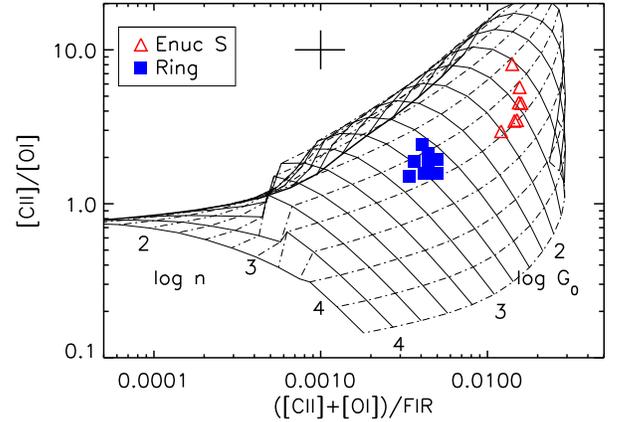}
\caption{Plot of [CII]$158\mu$m/[OI]$63\mu$m as a function of ([CII]$158\mu$m+[OI]$63\mu$m)/FIR luminosity ratios, emitted from a PDR according to a grid of $G_0$ and $n_H$, for the ring and Enuc S, as defined in Section \ref{heating}. For clouds illuminated on all sides by $G_0$, the observed ([O I]$ 63 \mu$m + [C II]$ 158 \mu$m)/FIR ratio is the ratio of the escaping intensities of line photons to grain continuum for each cloud. This corresponds to I([O I] 63 $\mu$m + [C II] 158 $\mu$m)/$2\times I_{FIR}$, where $I_{FIR} = 2 \times 1.3 \times 10^{-4} G_0$ ergs cm$^{-2}$ s$^{-1}$ sr$^{-1}$. Only [OI]$63\mu$m data from Enuc S and the ring are available. The error bars represent the average uncertainties derived from photometric errors.}
\label{wolfirefig}
\end{figure}

Using the neutral gas component of the [CII]$158\mu$m flux and the [OI]$63\mu$m flux for the ring and Enuc S, we estimate $G_0$ and $n_H$ throughout these regions. We use the notation that $G_0$ is measured in units of the Habing radiation field \citep{habing68}. We first assume that, after subtracting the ionized component, all the remaining [CII]$158\mu$m emission and all [OI]$63\mu$m and FIR flux is emitted from PDRs irradiated by massive stellar clusters. Following \citet{kaufman06}, the grid in Figure \ref{wolfirefig} shows a variation of the incident radiation field $G_0$ between $\sim80$ in the Enuc S region and $800$ in the ring. The gas density is $10^{3-3.5}$ cm$^{-3}$ in both regions. These values are similar to local star-forming spiral galaxies such as NGC 695 and NGC 3620 \citep{malhotra01}.
The density is a factor of 10 lower than in M82, but the value of $G_0$ in the ring is comparable to M82 \citep{kaufman99}. The larger $G_0$ in the ring presumably signals an increased ionization parameter, which measures the number of ionizing photons per hydrogen atom. 
This is consistent with our estimate above that $\sim33\%$ of the [CII]$158\mu$m originates from ionized regions. 

However, when considering detailed fits of the dust SED of the ring instead of just the FIR emission line ratios, a different picture of the intensity of the ionizing field emerges.  Using \citet{draine07} models, \citet{aniano11} estimate that only 25\% of the TIR is being emitted from regions where $G_0 > 10^2$.  For Enuc S this fraction is only about 8\%. Therefore, if we use these estimates for the infrared emission together with the PDR models of \citet{kaufman06}, we predict a lower ionizing radiation field intensity for the ring, much closer to a value of $G_0 \sim 10^{2.3}$, with a corresponding $n_H\sim 10^{3.5}$ cm$^{-3}$. These significantly different estimates for the $G_0$ and $n_H$ in the ring PDRs have important implications for the production of the warm $H_2$ emission we measure in the IRS data, and we will carry both of these estimates forward when discussing the source of the $H_2$ emission in Section 4.3. For Enuc S \citet{aniano11} predict a PDR fraction of the TIR flux of $\sim8\%$ in Enuc S. This will make ($F$([CII])+$F$([OI]))/$FIR>0.1$ in Enuc S, out of the range modeled by \citet{kaufman06} for PDR regions. All the line emission could be coming from the diffuse ISM and not PDRs in Enuc S and therefore we cannot predict the $G_0$ and $n_H$.

Recently using Herschel/PACS to observe a number of infrared starburst galaxies, \citet{gracia11} established a correlation between [CII]$158\mu$m/FIR and regions with high star formation efficiency, $L_{FIR}/M_{H_2}$, where [CII]$158\mu$m/FIR drops only when $L_{FIR}/M_{H_2}> 80 L_{\odot}/M_{\odot}$. 
Based on PACS $70\mu$m and $100\mu$m fluxes from \citet{sandstrom10} and CO 2--1 fluxes from \citet{hsieh08} (assuming $X_{CO}=3.0\times10^{20}$cm$^{Ð2}$ (K km s$^{Ð1}$)$^{Ð1}$, and $R_{21}\sim1.3$), we calculate $L_{FIR}/M_{H_2}\sim50 L_{\odot}/M_{\odot}$ in the ring of NGC 1097, lower than the value of $L_{FIR}/M_{H_2}$ for which [CII]$158\mu$m/FIR decreases for the luminous starbursts in \citet{gracia11}. Although [CII]$158\mu$m/FIR is lower in the ring than Enucs, it is consistent with other starburst galaxies having similar values of $L_{FIR}/M_{H_2}$.



\subsection{Gas heating efficiency and PAH abundance}
\label{discgas}

Theoretical studies have shown that the extent of grain charging, and therefore the efficiency of photoelectric heating, is not only dependent on environmental conditions, but also on grain size and grain species. \citet{bakes94} estimated that approximately half of the heating is from grains with less than about 1500 C-atoms ($\sim15 \AA$). 
Using IRAS and ISO observations of the [CII]$158\mu$m, [OI]$63\mu$m and ${\rm H_2}$ lines, \citet{habart04} quantified the photoelectric efficiencies attributed to the PAH, VSG (very small grains) and BG (big grains) populations in the $\rho$ Oph cloud, finding ${\rm\epsilon_{PAH}}= 3\%$, ${\rm\epsilon_{VSG}}=1\%$, and ${\rm\epsilon_{BG}}=0.1\%$ respectively. This suggests that PAHs are the most efficient gas heating channel in star-forming galaxies and therefore drive the heating of the neutral gas regions \citep{watson72,holl99}. Indeed, \citet{helou01} showed that the PAH contribution to the gas heating appears constant in galaxies with increasing $F(60\mu$m)/$F(100\mu$m), and dominant over the contribution of large grains, based on the relatively constant [CII]$158\mu$m/PAH against a rapidly falling [CII]$158\mu$m/FIR. This suggests that PAHs dominate the heating of the [CII]-emitting gas (via the photoelectric effect), while large grains dominate the FIR emission from galaxies \citep{malhotra01}. 
  
A decrease in the [CII]$158\mu$m relative to the FIR emission could also be caused by a decrease in the photoelectric gas heating efficiency associated with highly charged grains. While the [CII]$158\mu$m/PAH ($5.5 -14\mu$m) is lower in the ring than in the Enucs in NGC 1097, we find no correlation between [CII]$158\mu$m/PAH ($5.5 -14\mu$m) and the $11.3/7.7\mu$m PAH ratio, which measures PAH ionization. Although the average $11.3/7.7\mu$m PAH ratio is lower on average in the ring than in the Enucs, the regions in the Enucs with similar $11.3/7.7\mu$m PAH ratios have clearly higher [CII]$158\mu$m/PAH ($5.5 -14\mu$m). This means that the variation of [CII]$158\mu$m/PAH ($5.5 -14\mu$m) between the Enucs and the ring cannot be attributed solely to PAH ionization. 
However, we have seen in Figure~\ref{oipahfig} that [OI]$63\mu$m partially compensates for the decrease of [CII]$158\mu$m/PAH ($5.5 -14\mu$m) ratio in the ring. 
As seen in Section 3.3, ([CII]$158\mu$m+[OI]$63\mu$m)/PAH in Enuc S is $\sim1.5$ times higher than in the ring. After subtracting the ionized gas contribution to the [CII]158$\mu$m emission, the heating efficiency ([CII]158$\mu$m+[OI]63$\mu$m)/PAH in Enuc S is $\sim1.9$ times higher than in the ring. So by subtracting the ionized gas component, we obtain an estimate of the difference in the gas heating efficiency between the ring and Enuc S which is much larger than our previous estimates.

Since the [SiII]$34.8\mu$m line is also a major coolant we can include this emission in the overall energy budget of NGC 1097. 
Following \citet{kaufman06}, the [SiII]$34.8\mu$m flux emitted from PDRs increases relative to the [SiII]$34.8\mu$m flux emitted from HII regions as the electron density $n_e$ increases. The [SiII]$34.8\mu$m line can be as luminous as the [OI]$63\mu$m, and ([CII]+[OI]+[SiII])/PAH(5.5 -14$\mu$m) is similar in the ring and in Enuc S. However, this ratio would not represent accurately the gas heating efficiency, as it includes also the contributions to [CII]$158\mu$m and [SiII]$34.8\mu$m from the ionized gas. 
Therefore we can only estimate upper limits on the photoelectric heating efficiency of the PAHs in the PDRs using ([CII]$_{neut}$ + [OI] + [SiII])/PAH($5.5-14\mu$m) for the ring and Enuc S, respectively, with [CII]$_{neut}$ representing the neutral gas component of the [CII]$158\mu$m emission, calculated in Section \ref{pdr}. This results in a PAH photoelectric heating efficiency of $5.2\pm0.3\%$ in the ring and $7.7\pm0.8\%$. 
Thus we have seen that adding [SiII]$34.8\mu$m brings down the difference in photoelectric heating efficiency between the ring and Enuc S to 3$\sigma$ for ([CII]$_{neut}$ + [OI] + [SiII])/PAH($5.5-14\mu$m).

\begin{figure}
\epsscale{2.3}
\plottwo{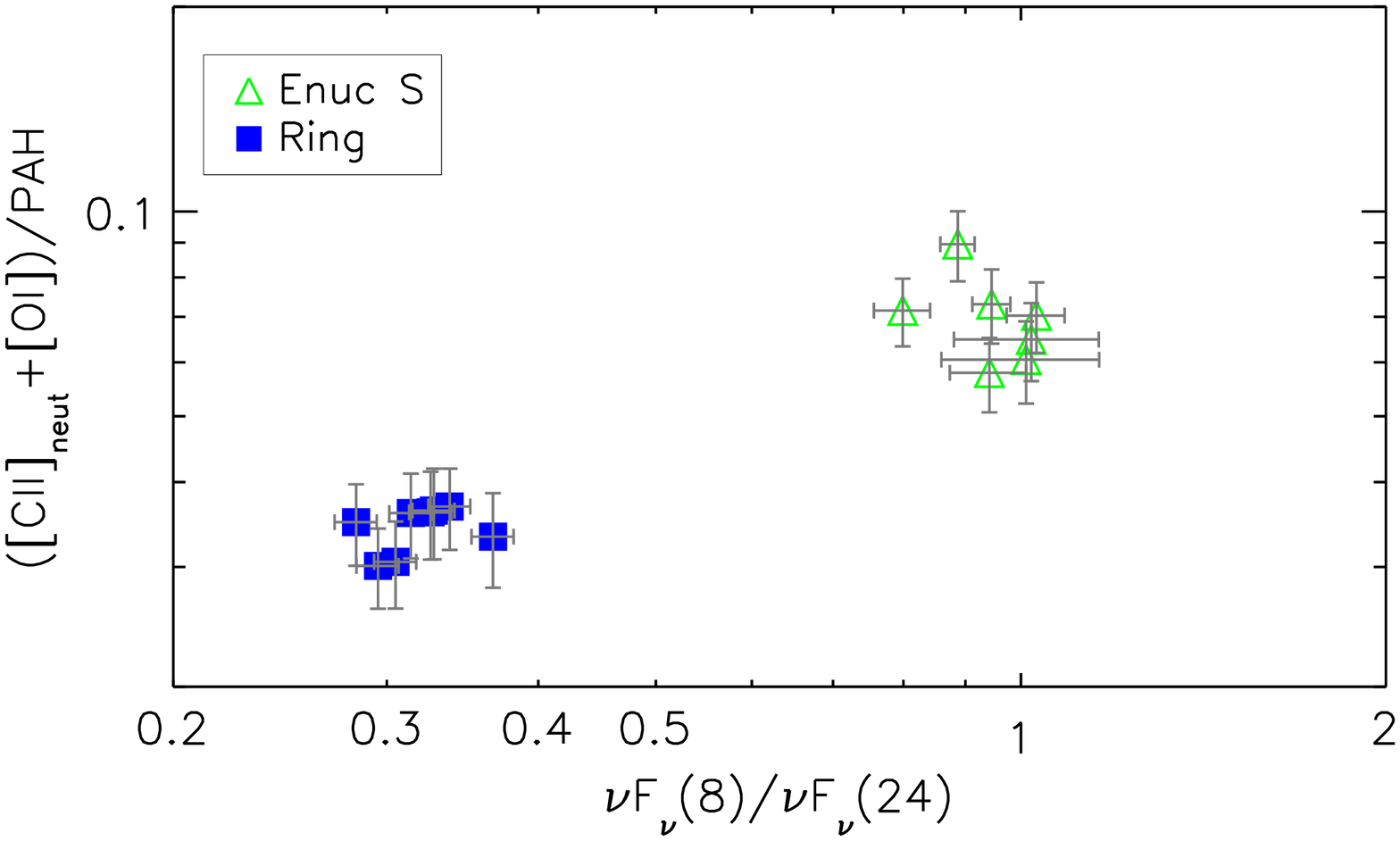}{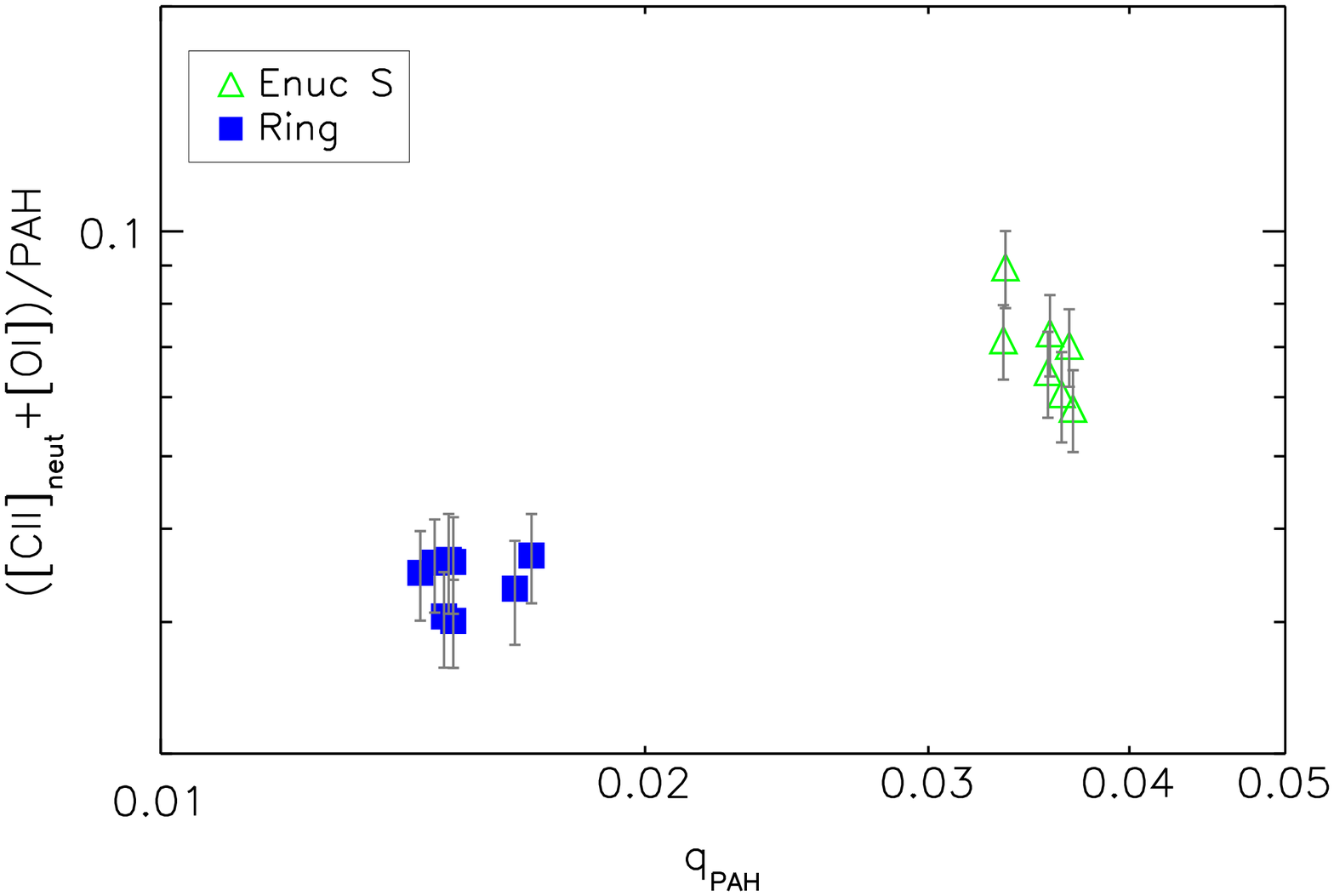}
\caption{Upper: Plot of the ([CII]$158\mu$m+[OI]$63\mu$m)/PAH ($5.5-14\mu$m) ratio as a function of F$_\nu(8\mu$m)/F$_\nu(24\mu$m) as traced by $Spitzer$-IRAC bands for the regions defined in Section \ref{heating}. Lower: Plot of the ([CII]$158\mu$m+[OI]$63\mu$m)/PAH ($5.5-14\mu$m) ratio as a function of the PAH mass abundance $q_{\rm PAH}$, calculated using the dust emission models of \citet{draine07}.}
\label{rubinfig}
\end{figure}

Using IRAC 8$\mu$m and MIPS $24\mu$m photometry of the Large Magellanic Cloud (LMC), \citet{rubin09} noted that the ratio between the [CII]$158\mu$m and the total infrared flux TIR decreases with decreasing F($8\mu$m)/F($24\mu$m), which they use as a measure of PAH/VSG flux in regions with bright PAH emission.
Variations of the $F_{\nu}(8\mu$m)/$F{_\nu}(24\mu$m) ratio arise mainly from the destruction of PAHs in HII regions and heating of VSGs to high temperatures in regions with intense radiation fields. If PAHs are destroyed, their emission in the infrared (at $8\mu$m) decreases relative to the infrared emission from other grains (at $24\mu$m). In regions with $G_0>100$, $F{_\nu}(24\mu$m)/$F_{TIR}$ will increase with starlight intensity, and the emission at $24\mu$m will increase relative to $8\mu$m. To verify if this is the case for NGC 1097, in Figure \ref{rubinfig} (upper) we present a plot of the ([CII]+[OI])/PAH($5.5 -14\mu$m) ratio for the ring and the Enuc N and S as a function of the $F_{\nu}(8\mu$m)/$F{_\nu}(24\mu$m) ratio, calculated from the IRAC $8\mu$m and MIPS $24\mu$m maps convolved to the PACS 160$\mu$m beam size \citep{aniano11}. The average $F_{\nu}(8\mu$m)/$F{_\nu}(24\mu$m) ratio in the ring in NGC 1097 is $\sim0.37\pm0.015$, the same found for locations in the LMC with the highest [CII]$158\mu$m surface brightness, whereas the average values of $F_{\nu}(8\mu$m)/$F{_\nu}(24\mu$m) for Enuc N and S are $\sim1.1\pm0.1$, which correspond to the average value observed in the LMC for similar [CII]$158\mu$m surface brightness. The ring has both lower ([CII]$158\mu$m+[OI]$63\mu$m)/PAH and $F_{\nu}(8\mu$m)/$F{_\nu}(24\mu$m) than the extranuclear regions. Therefore, the low ([CII]$_{neut}$+[OI])/PAH($5.5 -14\mu$m) and $F_{\nu}(8\mu$m)/$F{_\nu}(24\mu$m) might be due both to a lack of PAH grains, and higher starlight intensity. We should note that $F_{\nu}(8\mu$m)/$F{_\nu}(24\mu$m) is not an independent parameter, as it is strongly correlated with $F_{\nu}(60\mu$m)/$F{_\nu}(100\mu$m).

Using \citet{draine07} models we can calculate the PAH mass fraction, defined as the fraction of grains with 1000 C atoms or less to all dust grains, as a result of calculating infrared emission spectra for dust heated by starlight. In Figure \ref{rubinfig} (lower) we present a plot of the ([CII]$_{neut}$+[OI])/PAH($5.5 -14\mu$m) ratio for the ring and the Enuc N and S as a function of the PAH mass fraction $q_{\rm PAH}$, as measured in the $q_{\rm PAH}$ maps of NGC1097 constructed by \citet{aniano12}. The trend for ([CII]$_{neut}$+[OI])/PAH($5.5 -14\mu$m) as a function of $q_{\rm PAH}$ is similar to the trend seen for the $F_{\nu}(8\mu$m)/$F{_\nu}(24\mu$m) ratio, with $q_{\rm PAH}\sim0.018$ in the ring and $\sim0.035$ in the Enuc regions, a difference of a factor of 2. The median $q_{\rm PAH}$ in the SINGS sample is also 0.035 \citep{draine07b}. This suggests that the variation of the $F_{\nu}(8\mu$m)/$F{_\nu}(24\mu$m) between the ring and the Enucs in NGC 1097 is a consequence of the variation of the PAH abundance relative to other types of grains. As long as PAHs dominate photoelectric heating and therefore [CII]$158\mu$m and [OI]$63\mu$m cooling, the ([CII]$_{neut}$+[OI])/PAH($5.5 -14\mu$m) ratio should not change with PAH abundance. As seen in Figure \ref{rubinfig}, ([CII]$_{neut}$+[OI])/PAH($5.5 -14\mu$m) decreases with decreasing $q_{\rm PAH}$ and $F_{\nu}(8\mu$m)/$F{_\nu}(24\mu$m) ratio. This could be a consequence of an increase of the role of VSGs on the photoelectric gas heating, as the PAH mass fraction decreases and the VSG abundance increases faster than the PAH abundance. A significant portion of [CII]$158\mu$m and [OI]$63\mu$m could be heated by VSGs, increasing ([CII]$_{neut}$+[OI])/PAH($5.5 -14\mu$m). On the other hand, it may be that when $q_{\rm PAH}$ increases, the PAH size distribution shifts to smaller sizes, which may have higher photoelectric heating efficiencies.

Figure \ref{rubinfig} can also be interpreted by considering an alternative aspect of photoelectric heating efficiency. The efficiency of the photoelectric effect on a grain is traditionally defined
\citep[e.g.][]{holl99} as the ratio of gas heating
to the grain UV absorption rate. Observationally, this is often approximated
as the ratio of [CII]$158\mu$m or [CII]$158\mu$m+[OI]$63\mu$m to the infrared emission from dust,
assuming that [CII]$158\mu$m and [OI]$63\mu$m account for most of the cooling of the heated gas,
and total infrared emission accounts for most of the UV absorbed by grains.
However, photoelectric heating is dominated by the smallest grains in the ISM
\citep{watson72,jura76}, whereas the total infrared emanates primarily from
the larger grains. To avoid this complication,
\citet{helou01} used [CII]$158\mu$m/F(5-10$\mu$m) as an estimate of efficiency
limited to PAH, and showed that it varies considerably less
than [CII]$158\mu$m/FIR, thus demonstrating that the smallest grains, and more specifically
PAHs, do indeed dominate photoelectric heating. However, these definitions do not account for the shifting balance of the diverse dust species in mass ratio or absorption of heating radiation.
In Figure \ref{rubinfig} we see a clear difference in $q_{\rm PAH}$, the mass fraction of PAHs to large grains, between the starburst ring and Enuc S, much larger than the difference in ([CII]$158\mu$m + [OI]$63\mu$m)/PAH. A KS test revealed that the difference in $q_{\rm PAH}$ between the ring and Enuc S is significant. Using dust mass and $q_{\rm PAH}$ maps from \citet{aniano12}, constructed using \citep{draine07b} models, we are able to measure the total PAH mass in the ring and Enuc S and estimate the amount of [CII]$158\mu$m and [OI]$63\mu$m line emission per unit PAH mass in the ring and Enuc S.  The values for ([CII]$158\mu$m+[OI]$63\mu$m)/M$_{PAH}$ are $~70 L_{\odot}/M_{\odot}$ for enuc S and $~170 L_{\odot}/M_{\odot}$ for the ring, and may be interpreted as rough estimates of the amount of photoelectric heating contributed per unit mass of PAHs. These values are for PDR gas only, i. e., the ionized gas fraction of [CII]$158\mu$m was subtracted before deriving these quantities. These ratios suggest that per unit mass, PAHs in the ring are responsible for generating significantly more, by about a factor of two, [CII]$158\mu$m and [OI]$63\mu$m emission than PAHs in Enuc S, in spite of the slightly lower ([CII]$158\mu$m+[OI]$63\mu$m)/PAH flux ratio. In other words, the photoelectron production rate per unit mass of PAH is higher in the ring, even though the photoelectric heating efficiency is slightly lower. This difference in ratios is simply a reflection of the greater heating intensity available to the PAHs in the ring, making it possible for a PAH grain to process more photons per unit time, even if the photoelectric heating efficiency is lower, possibly because of increased ionization of the PAHs, or because of more frequent two-photon events. Since the PAHs are excited stochastically by one photon at a time, the increased production rate per unit mass can be seen as the result of approximately doubling the PAH excitation rate. This reflects how the mean starlight intensity $\langle U\rangle$ varies between the ring and Enuc S. Using $\langle U\rangle$ maps from \citet{aniano12}, constructed using \citep{draine07b} models, we measured that $\langle U\rangle$ in the ring is $\sim3.6$ times higher than in Enuc S, with an uncertainty of $\sim30\%$. Given the uncertainties involved, the ratio of $\langle U\rangle$ and the ratio of photoelectric production rate per unit mass between the ring and Enuc S are consistent.

\subsection{What is exciting the $H_2$?}
\label{h2section}

We have seen that the lower [CII]$158\mu$m/PAH observed in the ring is partially compensated by the increase of the [OI]$63\mu$m line, as it becomes a more efficient coolant in denser and warmer regions.
However, another mechanism that may be responsible for gas heating in these regions is shock heating. \citet{holl89} predicts fast ($>30$ kms$^{-1}$) J shocks to have strong [OI] emission. According to these models, the [OI]$63\mu$m line is typically stronger than the [CII]$158\mu$m line in regions where J-shock heating is important. However, [OI]$63\mu$m/[CII]$158\mu$m varies between $0.3 - 0.5$ in the ring \citep{beirao10}. Therefore fast shocks are not the dominant heating source of the ISM in the ring, at least not at scales probed by the PACS beam, although they could still be responsible for much of the [OI]$63\mu$m emission. Slow ($<30$ kms$^{-1}$) C shocks could still be present in the ring, as they emit only weak [OI]$63\mu$m \citep{holl89}.

By using the radiation field estimates from [CII]$158\mu$m and [OI]$63\mu$m, we can predict the $H_2$ emission from PDRs. Assuming the radiation field intensity $G_0$ calculated in section \ref{pdr}, the $H_2$ fluxes from the ring can be reproduced using a pure PDR model by \citet[e.g.]{kaufman06} with high density molecular gas ($n\sim10^5 cm^{-3}$). Such a molecular gas density is quite plausible, as \citet{kohno03} reported the existence of HCN (1-0) emission in the nucleus and ring, for which molecular gas densities of $\sim10^5$ cm$^{-3}$ are necessary \citep[e.g.][]{meijerink07}. For $G_0=10^{2.25}$ and $n_H=10^{3.5}$ calculated in Section 4.1, and assuming that the fraction of TIR flux from PDRs with these conditions is responsible for 25\% of the total TIR emission, we predict that such PDRs will emit a total of $H_2$ flux of $F(H_2 S(0)-S(3))\sim2.1\times10^{-16}$ Wm$^{-2}$. Our measured total $H_2$ flux is $F_{H_2}\sim1.3\times10^{-15}$ Wm$^{-2}$, about a factor of 6 above the modeled value. Therefore, if most of the dust emission is coming from low $\langle U\rangle$ diffuse ISM, it is very difficult to account for the observed $H_2$ from the ring with intense PDRs. The $H_2$ lines must predominately come from some other type of region.

In the Milky Way there is increasing evidence that the diffuse ISM has pockets of warm $H_2$ that radiate in the $H_2$ rotational lines, where the excited $H_2$ is in excess of what would be produced by UV-pumping. Both $H_2$ and PAH emission have been observed in gas clouds illuminated by starlight \citep[e.g.][]{habart04}.
\citet{ingalls11} used Spitzer-IRS to show that high-latitude translucent clouds, illuminated by just the local starlight background ($\langle U\rangle \sim1$), may contain small amounts of hot, collisionally excited $H_2$. In these clouds, $F(H_2 S(2))/F(PAH7.7) \sim 0.002$ on average. In the ring of NGC 1097, $F(H_2 S(2))$/F(PAH7.7)$ \sim 0.0015$, so translucent clouds with low incident radiation are a plausible origin for the observed $H_2$ emission in the ring.
The warm regions that produce the $H_2$ emission observed by \citet{ingalls11} will also radiate in [CII]158$\mu$m and [OI]63$\mu$m. Combining the $H_2$ measurements in \citet{ingalls11} with the [CII]$158\mu$m results of \citet{ingalls02} gives $F(H_2 S(0)-S(2))$/[CII] $\sim0.4-1$ for Milky Way translucent cloud positions detected in $H_2$ emission, which is much higher than the observed $F(H_2 S(0)-S(2))$/[CII]$\sim 0.090$ in the ring. \citet{ingalls11b} predict that the translucent clouds should have [OI]/$F(H_2 S(0)-S(2))< 0.19$, well below the observed [OI]/$F(H_2 S(0)-S(2))\sim 4.5$ in the ring. Thus, it seems possible that much of the $H_2$ rotational line emission from the ring could come form warm regions in the $H_2$ clouds, but these regions do not dominate the [CII]$158\mu$m or [OI]$63\mu$m emission we observe. 

\begin{figure}
\epsscale{1.2}
\plotone{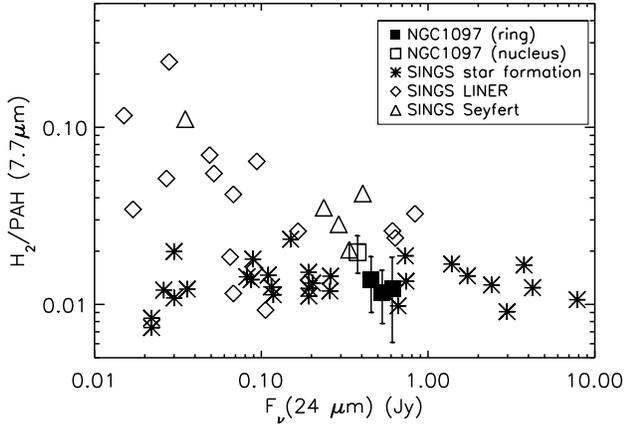}
\caption{Ratio of $H_2$ luminosity summed over the S(0)-S(3) lines to $7.7\mu$m PAH luminosity vs. $\nu F_{\nu}$(24 $\mu$m) continuum luminosity. This ratio indicates the relative importance of mechanical heating and star formation power. The open square is the average for the nucleus, and the full squares are the values for three regions in the ring shown in Figure \ref{h2fig}. All other points are ratios calculated for galaxies in the SINGS sample \citep{roussel07}.}
\label{oglefig}
\end{figure}

In Figure \ref{oglefig} we plot the ratio of $H_2$ luminosity in the 0-0 S(0)-S(3) lines over $L_{PAH7.7}$ versus $F_{24}$ for regions 1, 2 , and 3 in the ring and in the nucleus and compare these measurements with data from the SINGS sample of nearby galaxies \citep{roussel07}. While the values for the ring are similar to those of star forming galaxies \citep{roussel07}, which have a median ratio $F(H_2) S(0)-S(3)$/L(PAH7.7) = 0.011, the nucleus is slightly larger, with a value of $\sim0.02$, more consistent with the LINER and Seyfert nuclei in \citet{roussel07}, although there is a large scatter among the galaxies with AGN.

The nucleus of NGC 1097 has a weak AGN, apparently fed by an inflow of gas inside the ring \citep{prieto05}. It is possible therefore, that X-rays from the AGN might contribute to the heating of the $H_2$ gas in the immediate vicinity of the nucleus. The cooling by $H_2$ rotational lines in X-ray-dominated regions (XDRs) is 2\% of the total gas cooling \citep{maloney96} for a temperature of 200 K, typical of warm $H_2$ that emits in the mid-infrared. At this temperature, the ratio of the first four rotational lines to the total rotational line luminosity is $L(H_2 S(0)-S(3))/L(H_2)= 0.58$. Combining the above factors, we estimate a maximum $H_2$ to X-ray luminosity ratio of $L(H_2 S(0)-S(3))/L_X(2-10$ keV)$= 0.01$. This ratio is conservative, since it assumes that all of the X-ray flux from the AGN is absorbed by the XDR.
The nucleus of NGC 1097 has an X-ray luminosity of $L_X (2-10$ keV$)=1.73\times10^{-15}$ W m$^{-2}$ \citep{lutz04}. Therefore the maximum $H_2$ luminosity should be $L(H_2)\sim1.7\times10^{-17}$ W m$^{-2}$, significantly lower than the measured $L(H_2 S(0)-S(3))$ in the nucleus, which is $\sim2.3\times10^{-16}$ W m$^{-2}$. Considering L($H_2$ S(0)-S(3))/L($H_2$)= 0.58, the contribution of X-rays for heating the $H_2$ is less than 5\%. This is consistent with the near infrared rotational-vibrational spectra taken across the nucleus by \citet{reunanen02}.

Another possible source for $H_2$ heating are cosmic rays from supernovae. The energy released by the dissociation of each $H_2$ molecule by cosmic rays is $\sim$12 eV, including the contribution from $H_3^+$ recombination and $H_2$ re-formation \citep[e.g.][]{lepetit06}. To balance line cooling, the ionization rate per $H_2$, $\zeta_{\rm H2}$, must be few $\times$ $10^{-14}~{\rm s^{-1}}$. For such a rate, cosmic rays are the main destruction path of $H_2$ molecules.
From the Table~\ref{h2tab} the total $H_2$ emission-line luminosity is $5.91 \times 10^{40} ~ {\rm erg ~ s^{-1}}$ for the ring and $1.02 \times 10^{40} ~ {\rm erg ~ s^{-1}}$ for the nucleus. Dividing this luminosity by the warm $H_2$ masses of $1.52 \times 10^8 ~ M_\odot $ and $9.6 \times 10^6 ~ M_\odot $ respectively, we can estimate a cooling rate through the $H_2$ lines of $3.2 \times 10^{-25} ~ {\rm erg ~ s^{-1} ~ H_2^{-1}}$ and $8.8 \times 10^{-25} ~ {\rm erg ~ s^{-1} ~ H_2^{-1}}$ respectively. 
The mass fraction of the gas in the molecular state depends on the rate of ionization and gas density $\zeta_{\rm H2} /n_{\rm H2}$. Dividing the $H_2$ S(0)-S(3) luminosity by the energy released per ionization, we estimate the required ionization rate, $\zeta_{\rm H_2} \sim 1 \times 10^{-13}~{\rm s^{-1}}$ for the ring and $\zeta_{\rm H_2} \sim 6 \times 10^{-13}~{\rm s^{-1}}$ for the nucleus. This value is larger than what is inferred from $H_3^+$ observations in the Milky Way, by a factor $200-1000$ for the diffuse interstellar medium \citep{indirolo07}, and by more than a factor 20 for the molecular gas within 200 pc from the Galactic center \citep{goto08}. It is therefore unlikely that the warm molecular gas in the ring and nucleus of NGC 1097 is heated solely by cosmic rays.

\section{Summary}

In this paper we have attempted to understand the energetics and the physical conditions in the ISM in NGC 1097, using Herschel-PACS and Spitzer-IRS. 
Our goal was to study how heating and cooling of the ISM vary with the ISM phases, studying the origin of the variations of heating efficiency with dust temperature, and the properties of the warm $H_2$ in regions with different ISM heating mechanisms. 
We focused on the central region, where a ring of star formation and an AGN are located, and in two extranuclear regions at the ends of the bar, which we named Enuc N and S. 
We divided the ring, Enuc N and Enuc S regions into $12\arcsec\times12\arcsec$ regions. 
After subtracting the fraction of [CII]$158\mu$m flux emitted from ionized gas, estimated as $\sim33\%$ in the ring and $\sim5\%$ in enuc S, we examined the correlations between the [CII]$158\mu$m/PAH ratio and the $70\mu$m/100$\mu$m, the PAH ratio $11.3/7.7\mu$m. We implemented PDR models by \citet{kaufman06}, and used the [CII]$158\mu$m/[OI]$63\mu$m ratio and ([CII]$158\mu$m+[OI]$63\mu$m)/FIR to plot a grid of the local FUV flux $G_0$, and the PDR density $n$ and diagnose the properties of the ISM on the ring and Enuc S.
We also calculated the total $H_2$ and the PAH $7.7\mu$m fluxes in three regions in the ring and one centered in the nucleus to determine the origin of the warm $H_2$ excitation. As a result of our analysis we find:

\begin{itemize}
\item Assuming that PDRs account for all [OI]63um, FIR flux and 67\% of the [CII]$158\mu$m, the average intensity of the radiation field $G_0$ is $\sim80$ in Enuc S and $\sim800$ in the ring. The gas density is $10^{3-3.5}$ cm$^{-3}$ in both regions. If instead 75\% of the FIR flux in the ring arises from low starlight intensity regions and only 25\% of the FIR flux arises from PDRs, then $G_0$ is $\sim10^{2.3}$ and $n_H \sim 10^{3.5}$ in the ring.
\item The ratio [CII]$158\mu$m/PAH($5.5 - 14\mu$m) is a factor of $\sim1.7$ lower in the ring than in the Enuc regions and [CII]$158\mu$m/FIR is a factor of $\sim4$ lower in the ring than at the southern edge of the bar. The average $11.3/7.7\mu$m PAH ratio is lower in the ring than at the ends of the bar, suggesting a larger fraction of ionized grains in the starburst ring. However, no clear correlation was seen between the $11.3/7.7\mu$m PAH ratio and the [CII]$158\mu$m/PAH($5.5 - 14\mu$m) ratio across the complete dataset, implying that grain ionization is not the sole driver of the grain heating efficiency variations. The ratio ([CII]$158\mu$m+[OI]$63\mu$m)/PAH is still lower in the ring than the ends of the bar, although the difference is much smaller and the scatter larger than in [CII]$158\mu$m/PAH($5.5 - 14\mu$m).
After the removal of the ionized gas component of the [CII]$158\mu$m emission and adding [SiII]$34.5\mu$m, we calculated a gas heating efficiency in the ring of $\sim5.2\%$ and $\sim7.7\%$ in Enuc S. 
The amount of [CII]$158\mu$m and [OI]$63\mu$m line emission per unit PAH mass is roughly a factor of two larger in the ring than in Enuc S.  
Since the PAHs are excited stochastically by photons, the increased production rate per unit mass can be seen as the result of approximately doubling the PAH excitation event rate.
\item 
Using the [CII]$158\mu$m and [OI]$63\mu$m emission lines and total FIR emission from the ring are used in conjunction with standard PDR models, the derived radiation field can account for nearly all the warm $H_2$ emission measured with Spitzer/IRS. However, if we incorporate the results of the dust SED modeling, which suggests nearly 75\% of the FIR flux comes from dust heated by low starlight intensities in the diffuse ISM, there is a large excess of $H_2$ which cannot be excited in PDRs. In analogy to what is seen in the ISM of the Milky Way \citep{ingalls11}, we suggest this $H_2$ emission in the NGC 1097 ring may be coming from pockets of warm $H_2$ gas inside translucent molecular clouds in the diffuse ISM.  
\item The $H_2$ S(0)-S(3)/PAH7.7$\mu$m ratio is enhanced in the nucleus (the site of a weak AGN) relative to the ring by a factor of two, as is the fraction of warm ($T \sim 400$ K) molecular gas. X-rays and cosmic rays cannot provide more than 1-5\% of the energy necessary to produce the $H_2$ emission in the nucleus. Shock waves, possibly associated with the central AGN, may be an important contributor to gas heating in the nucleus. 
\end{itemize}

This paper is an example of the study that can be carried out by carefully combining Herschel/PACS and Spitzer/IRS data focusing on starburst rings or other resolved regions of enhanced star-formation in nearby galaxies.  In future papers we will explore the energy balance in the ISM of local star-forming galaxies within the KINGFISH sample, and apply these methods of linking the atomic, and molecular gas feature emission with the dust properties, to more luminous starburst galaxies observed with Spitzer and Herschel.

\begin{acknowledgements}

We would like to thank Gregory Brunner and Sebastian Haan for the code used to construct the Spitzer-IRS maps. We would like to thank Dario Fadda and Jeff Jacobson for software support.
This work is partially based on observations made with \textit{Herschel}, a European Space Agency Cornerstone Mission with significant participation by NASA. Support for this work was provided by NASA through an award issued by JPL/Caltech. PACS has been developed by a consortium of institutes led by MPE (Germany) and including UVIE (Austria); KU Leuven, CSL, IMEC (Belgium); CEA, LAM (France); MPIA (Germany); INAF-IFSI/OAA/OAP/OAT, LENS, SISSA (Italy); IAC (Spain). This development has been supported by the funding agencies BMVIT (Austria), ESA-PRODEX (Belgium), CEA/CNES (France), DLR (Germany), ASI/INAF (Italy), and CICYT/MCYT (Spain). Data presented in this paper were analyzed using ÒThe \textit{Herschel} Interactive Processing Environment (HIPE),Ó a joint development by the \textit{Herschel} Science Ground Segment Consortium, consisting of ESA, the NASA \textit{Herschel} Science Center, and the HIFI, PACS and SPIRE consortia.

\end{acknowledgements}

\newpage

\begin{sidewaystable}
\tiny
\centering
\caption{Emission line fluxes from Spitzer-IRS and PACS spectra (in units of $10^{-17}$ Wm$^{-2}$)}
\begin{tabular}{lccccccccccccccccc}
\hline
\hline
Region & Aperture Size & PAH &[ArII] & PAH & PAH & [ArIII] & [SIV] & PAH & PAH & [NeII] & [NeIII] & [SIII] & [SIII] & [SiII]& PAH total & PAH ratio & PAH ratio\\
 & & 6.2 $\mu$m& $6.99\mu$m& 7.7$\mu$m&8.6$\mu$m & $8.99\mu$m & $10.5\mu$m & 11.3$\mu$m& 12.6$\mu$m& $12.8\mu$m & $15.6\mu$m & $18.7\mu$m & $33.5\mu$m & $34.8\mu$m & $6.2-12.6\mu$m& 11.3/7.7& 6.2/7.7\\
& arcsec$^2$& & 15.8 eV & & & 27.6 eV & 34.8 eV & & & 21.6 eV & 41.0 eV & 23.3 eV & 23.3 eV & 8.2 eV & & &\\
\hline
 Nucleus &  96.8 & 2.44          & 0.103            & 11.2           & 1.93            & \nodata &    0.275         & 3.69        & 2.34               & 0.364           & 0.415              & 0.136           & 0.246           & 0.576              & 21.6 & 0.33 & 0.23\\
                 &         & $\pm$0.19 & $\pm$0.025 & $\pm$0.6 & $\pm$0.12 &\nodata & $\pm$0.003 & $\pm$0.07 &  $\pm$0.07 & $\pm$0.004 & $\pm$0.050 & $\pm$0.001 & $\pm$0.003 & $\pm$0.003 &\nodata &\nodata &\nodata \\
 Ring &  680.1 & 32.6        & 1.726           & 130.4      & 23.3           & 0.518             & 0.206            & 30.5         & 20.3          &4.59           & 0.205             & 1.18                & 1.72          & 3.81 & 237.1 & 0.23 & 0.25  \\
             &         & $\pm$1.3 & $\pm$3.68 & $\pm$4.6 & $\pm$1.0 & $\pm$0.013 & $\pm$0.008 & $\pm$0.5 & $\pm$0.7 & $\pm$0.01 & $\pm$0.003 & $\pm$0.01 & $\pm$0.01 & $\pm$0.01 & & & \\
 Enuc N & 1067.2& 1.68          & 0.048               & 5.79           & \nodata & 0.009              & 1.03         & 1.32              & 0.129 & \nodata &\nodata& \nodata & \nodata& \nodata &10.5 & 0.23 & 0.29  \\
                           & & $\pm$0.09 & $\pm$0.009 & $\pm$0.08 & \nodata & $\pm$0.002 & $\pm$0.15 & $\pm$0.11 & $\pm$0.002 & \nodata & \nodata&\nodata& \nodata & \nodata\\
Enuc S & 739.3& 1.59 & 0.074& 5.26 & 1.06& \nodata & \nodata& 1.32 & 0.66& 0.178 & \nodata &\nodata  & \nodata&  \nodata& 9.89 & 0.25 & 0.30 \\
 & & $\pm$0.09 & $\pm$0.002 & $\pm$0.07 & $\pm$0.17 & \nodata & \nodata & $\pm$0.10 & $\pm$0.06 & $\pm$0.35  & \nodata & \nodata&\nodata &\nodata  & \nodata\\ 
  \hline
\end{tabular}
\label{iontab}
\end{sidewaystable}

\begin{table}
\centering
\caption{$H_2$ line fluxes\tablenotemark{a} from low and high-resolution spectra}
\begin{tabular}{lccccc}
\hline 
\hline
Region & Area ($arcsec^2$) & S(0)$28.2\mu$m & S(1)$17.0\mu$m & S(2)$12.3\mu$m  & S(3)$9.7\mu$m \\
\hline
Nucleus & 96.8 & 0.55$\pm$0.13 & 6.13$\pm$0.78 & 6.21$\pm$0.79 &	10.1$\pm$1.01\\
Ring & 680.1 & 10.7$\pm$0.79 & 35.9$\pm$1.2 & 20.7$\pm$1.4 & 58.5$\pm$16.1 \\
Ring 1 &  96.8&	0.77$\pm$0.28 &	4.00$\pm$0.63 &	6.13$\pm$0.78 &	11.7$\pm$1.1 \\
Ring 2 &96.8	& 1.96$\pm$0.44 &	4.26$\pm$0.65 & 4.53$\pm$0.67 & 4.66$\pm$0.68  \\
Ring 3 &96.8	& 0.99$\pm$0.31 &	4.44$\pm$0.67 & 5.86$\pm$0.77 & 7.17$\pm$0.85 \\
Enuc N\tablenotemark{b} & 1067.2& 0.209$\pm$0.039 & \nodata & 0.04$\pm$0.01 & 0.561$\pm$0.010   \\
Enuc S\tablenotemark{b} & 739.3& 0.322$\pm$0.074 & \nodata & \nodata & 0.770$\pm$0.015 \\
\hline
\end{tabular}
\label{h2fluxtab}
\tablenotetext{1}{In units of $10^{-17}$ Wm$^{-2}$}
\tablenotetext{2}{From high-resolution spectra}
\end{table}


\begin{table}
\tiny
\centering
\caption{PAH band and line fluxes\tablenotemark{a} for each 12$\arcsec$ region}
\begin{tabular}{lccccccccc}
\hline \hline
Regions & PAH 7.7$\mu$m& PAH 11.3$\mu$m & PAH $6.2-12.6\mu$m &[SIII]$18.7\mu$m\tablenotemark{b} & [SIII]$33.5\mu$m\tablenotemark{b} & [SiII]$34.8\mu$m\tablenotemark{b} & [OI]$63\mu$m & [CII]$158\mu$m & FIR\tablenotemark{c} \\
\hline
\multicolumn{3}{l}{Ring} \\
1& 177$\pm$8& 42.9$\pm$0.7&	366$\pm$7& 2.02$\pm$0.09 & 4.15$\pm$0.01 &	7.19$\pm$0.02&	4.38$\pm$0.45 & 9.91$\pm$0.03 &2280\\
2&256$\pm$9 &63.9$\pm$1.0 &	545$\pm$5& 2.61$\pm$0.10 & 4.50$\pm$0.01 &	8.52$\pm$0.02&	7.37$\pm$0.21 & 17.4$\pm$0.1 &3363\\
3&195$\pm$6 &50.0$\pm$1.0 &	423$\pm$6& 2.01$\pm$0.11 & 3.95$\pm$0.01 &	7.46$\pm$0.02&	6.06$\pm$0.18 &14.2$\pm$0.1&2467\\
4&244$\pm$7 &59.7$\pm$1.1 &	514$\pm$7& 2.22$\pm$0.10 & 4.34$\pm$0.01&	8.23$\pm$0.02&	5.44$\pm$0.45& 15.3$\pm$0.1&3174\\
5&235$\pm$7 &60.6$\pm$1.0 &	508$\pm$7& 2.30$\pm$0.10 & 4.00$\pm$0.01 &	8.30$\pm$0.02&	6.64$\pm$0.35& 17.4$\pm$0.1&3317\\
6&171$\pm$5 &43.7$\pm$0.8 &	358$\pm$6& 1.45$\pm$0.14 & 3.00$\pm$0.01& 	6.88$\pm$0.01&	3.46$\pm$0.27& 12.6$\pm$0.1&2215\\
7&226$\pm$8 &57.3$\pm$1.1 &	474$\pm$9& 1.92$\pm$0.13 & 3.63$\pm$0.01 &	7.99$\pm$0.02&	5.49$\pm$0.55& 17.5$\pm$0.1&3101\\
8&172$\pm$8 &44.1$\pm$0.8 &	360$\pm$6& 1.68$\pm$0.13 & 3.05$\pm$0.01 &	7.14$\pm$0.02&	4.42$\pm$0.45& 12.8$\pm$0.1&2318\\
\multicolumn{3}{l}{Enuc N} \\
1&7.30$\pm$0.35 &2.02$\pm$0.05 &	15.9$\pm$0.3& 0.359$\pm$0.007&0.672$\pm$0.107 &0.629$\pm$0.038 &  \nodata&0.920$\pm$0.046& 71.4\\
2&6.73$\pm$0.43 &1.92$\pm$0.05 &	14.7$\pm$0.3& 0.291$\pm$0.006& 0.642$\pm$0.102& 0.601$\pm$0.036&  \nodata &0.712$\pm$0.035& 56.9\\
3&7.40$\pm$0.38 &2.16$\pm$0.06 &	15.1$\pm$0.3& 0.221$\pm$0.005& 0.526$\pm$0.084&0.486$\pm$0.015& \nodata &0.801$\pm$0.040& 75.4\\
4&7.54$\pm$0.39 &2.05$\pm$0.04 &	17.0$\pm$0.4& 0.232$\pm$0.005&0.703$\pm$0.111 & 0.655$\pm$0.039&  \nodata&0.889$\pm$0.044& 71.0\\
5&6.71$\pm$0.52&1.73$\pm$0.06 &	15.7$\pm$0.4& 0.117$\pm$0.003 &0.573$\pm$0.091&0.535$\pm$0.017 & \nodata &0.749$\pm$0.037& 56.9\\
6&9.75$\pm$0.28 &2.94$\pm$0.05 &	22.0$\pm$0.3& \nodata  &0.613$\pm$0.018 & 0.656$\pm$0.104 & \nodata & 1.01$\pm$0.05& 91.3\\
7&6.99$\pm$0.45 &2.05$\pm$0.06 &	16.2$\pm$0.4& \nodata &0.541$\pm$0.032&0.582$\pm$0.093 & \nodata&0.819$\pm$0.040& 71.2\\
8&8.43$\pm$0.56 &3.01$\pm$0.06 &	18.2$\pm$0.3& \nodata & \nodata &  \nodata&  \nodata&0.936$\pm$0.042& 103\\
9&8.86$\pm$0.56 &2.86$\pm$0.06 &	21.2$\pm$0.4& \nodata& 0.713$\pm$0.113&0.671$\pm$0.040 &  \nodata&1.11$\pm$0.05& 97.3\\
10&5.32$\pm$0.69 &1.67$\pm$0.09 & 13.4$\pm$0.3&  \nodata& 0.458$\pm$0.073&0.427$\pm$0.026 &  \nodata&0.676$\pm$0.033& 54.1\\
\multicolumn{3}{l}{Enuc S} \\  
1 & 8.96$\pm$0.48 & 2.65$\pm$0.04 & 20.7$\pm$0.4& \nodata	&	 \nodata & \nodata	 &0.360$\pm$0.102& 1.12$\pm$0.03  &118\\
2 & 10.4$\pm$0.5&2.82$\pm$0.04 &	23.7$\pm$0.4& \nodata &0.068$\pm$0.01 &	0.315$\pm$0.09 &	0.356$\pm$0.046& 1.31$\pm$0.01  &104\\
3  &11.6$\pm$0.5 &3.18$\pm$0.04 & 26.5$\pm$0.3& \nodata	& \nodata & \nodata	&	0.512$\pm$0.058& 1.86$\pm$0.02  &148\\
4  & 10.3$\pm$0.5&2.60$\pm$0.04 & 23.0$\pm$0.3& 0.058$\pm$0.024 &0.071$\pm$0.023 &	0.229$\pm$0.071&	0.232$\pm$0.036& 1.10$\pm$0.01  &	78.2\\
5 &13.0$\pm$0.6 &3.24$\pm$0.04 & 28.8$\pm$0.3&	0.068$\pm$0.032 & 0.110$\pm$0.032 & 0.436$\pm$0.131&	0.364$\pm$0.036& 1.74$\pm$0.01  &	 122\\
6 &5.45$\pm$0.51 &1.50$\pm$0.05 & 12.7$\pm$0.5&      \nodata            & \nodata	              &   \nodata      & 0.081$\pm$0.013 &   0.688$\pm$0.007   &53.4\\
7   &7.50$\pm$0.60 &1.93$\pm$0.05 &17.0$\pm$0.5&  \nodata& 0.056$\pm$0.02 &	0.305$\pm$0.108&	0.157$\pm$0.020&  0.945$\pm$0.010  	&67.2\\
\hline
\end{tabular}
\tablenotetext{1}{In units of $10^{-16}$ Wm$^{-2}$}
\tablenotetext{2}{This line was measured from SH ([SIII]$18.7\mu$m) and LH maps ([SIII]$33.5\mu$m and [SiII]$34.8\mu$m). The SH maps have an area of 1164 $arcsec^2$ for the ring, 684 $arcsec^2$ for Enuc N and 337 $arcsec^2$ for Enuc S. The LH maps have an area of  1392 $arcsec^2$ for the ring, 2188 $arcsec^2$ for Enuc N and $1193arcsec^2$ for Enuc S.}
\tablenotetext{3}{The FIR flux was calculated as FIR=TIR/2, with TIR calculated from \citet{dale02} using Spitzer MIPS $24\mu$m and Herschel PACS $70\mu$m and $160\mu$m}
\label{regiontab}
\end{table}

\begin{table}
\centering
\caption{$H_2$ Model Parameters}
\begin{tabular}{lcccccc}
\hline \hline
Regions & Area ($arcsec^2$) & T(K) & Ortho/Para & $N(H_2)$\tablenotemark{a} & $M(H_2)$\tablenotemark{b} & $L(H_2)$\tablenotemark{c}\\
\hline
Nucleus & 96.8 & 100 &  1.587 & 8.71$\pm$3.76 & 9.14$\pm$3.94 & 0.06  \\
& & 412$\pm$19 & 3.000 & 0.46$\pm$0.08 & 0.48$\pm$0.09 & 2.58 \\
Ring total & 680.1 & 119$\pm$11 & 1.972 & 16.0$\pm$5.0 &  151.0$\pm$47.2 &  2.75 \\
 & & 467$\pm$79 &  2.999 & 0.15$\pm$0.08 &  1.44$\pm$0.72 & 12.6  \\
Region 1 & 96.8 & 100 & 1.587 & 14.1$\pm$7.1 &  14.8$\pm$7.48 &  0.09  \\
& & 470$\pm$27 & 3.000 & 0.29$\pm$0.06 & 0.30$\pm$0.06 & 2.73 \\
Region 2 & 96.8 & 100 &  1.587 & 41.3$\pm$10.8 &  43.4$\pm$11.3 & 0.28 \\
& &  375$\pm$23 & 2.995 & 0.40$\pm$0.11 & 0.42$\pm$0.11 &  1.52  \\
Region 3 & 96.8 & 100 &  1.587 & 18.0$\pm$7.8 & 18.9$\pm$8.2  & 0.12  \\
& &  401$\pm$22 & 2.998 &  0.41$\pm$0.09 &  0.43$\pm$0.10 & 2.05  \\
\hline
\end{tabular}
\label{h2tab}
\tablenotetext{1}{In units of $10^{20}$ cm$^{-2}$}
\tablenotetext{2}{In units of $10^6 M_{\odot}$}
\tablenotetext{3}{In units of $10^6 L_{\odot}$}
\end{table}

\end{document}